\documentclass[11pt]{article}
\usepackage[hmargin=1in,vmargin=1in]{geometry}
\usepackage{paper}
\usepackage{thm-restate}
\usepackage{cleveref}
\usepackage{comment}
\usepackage{palatino}
\mathcode`l="8000
\begingroup
\makeatletter
\lccode`\~=`\l
\DeclareMathSymbol{\lsb@l}{\mathalpha}{letters}{`l}
\lowercase{\gdef~{\ifnum\the\mathgroup=\m@ne \ell \else \lsb@l \fi}}%
\endgroup

\newtheorem{theorem}{Theorem}[section]
\newtheorem{lemma}[theorem]{Lemma}
\newtheorem{claim}[theorem]{Claim}
\newtheorem{observation}[theorem]{Observation}

\newtheorem{remark}[theorem]{Remark}
\newtheorem{corollary}[theorem]{Corollary}

\newtheorem{definition}[theorem]{Definition}
\newtheorem{question}[theorem]{Question}

\def\eps{\e}

\newcommand{\lca}{\mathcal{LCA}}

\newcommand{\cL}{\mathcal{L}}
\newcommand{\bB}{\mathbb{B}}
\newcommand{\cM}{\mathcal{M}}
\newcommand{\bR}{\mathbb{R}}
\newcommand{\bM}{\mathbb{M}}

\newcommand{\trees}{\textsc{Tree}}
\newcommand{\planar}{\textsc{Planar}}
\newcommand{\polydom}{\textsc{PolyDom}}
\newcommand{\terrain}{\textsc{Terrain}}
\newcommand{\polysurf}{\textsc{PolySurf}}
\newcommand{\comb}{\textsc{Comb}}

\title{Spanners in Planar Domains 
    via Steiner Spanners and non-Steiner Tree Covers}
    \author{Anonymous}
\author{Sujoy Bhore \thanks{Department of Computer Science \& Engineering, Indian Institute of Technology Bombay. \href{}{sujoy@cse.iitb.ac.in}}\and 
	Balázs Keszegh \thanks{HUN-REN Alfréd Rényi Institute of Mathematics and ELTE Eötvös Loránd University, Budapest, Hungary.  \href{}{keszegh@renyi.hu}}\and
        Andrey Kupavskii \thanks{Moscow Institute of Physics and Technology and St. Petersburg State University, Russia, \href{}{kupavskii@ya.ru}} \and 
        Hung Le\thanks{University of Massachusetts Amherst, \href{}{hungle@cs.umass.edu}}\and
        Alexandre Louvet \thanks{LIPN, Université Sorbonne Paris Nord, \href{}{alexandre.louvet@mailo.fr}} \and  
         Dömötör Pálvölgyi \thanks{ELTE Eötvös Loránd University and HUN-REN Alfréd Rényi Institute of Mathematics, Budapest, \href{}{domotor.palvolgyi@ttk.elte.hu}}\and
	Csaba D. T\'oth\thanks{California State University Northridge, Los Angeles, CA, and Tufts University, Medford, MA, \href{}{csaba.toth@csun.edu}}
}

\date{}
\begin{document}

\maketitle

\begin{abstract}
We study spanners in planar domains, including polygonal domains, polyhedral terrain, and planar metrics. Previous work showed that for any constant $\eps\in (0,1)$, one could construct a $(2+\eps)$-spanner with $O(n\log(n))$ edges (SICOMP 2019), and there is a lower bound of $\Omega(n^2)$ edges for any $(2-\eps)$-spanner (SoCG 2015). The main open question is whether  
 a linear number of edges suffices and the stretch can be reduced to $2$. We resolve this problem by showing that for stretch $2$, one needs $\Omega(n\log n)$ edges, and for stretch $2+\eps$ for any fixed $\eps \in (0,1)$, $O(n)$ edges are sufficient. Our lower bound is the first super-linear lower bound for stretch $2$. 

En route to achieve our result, we introduce the problem of constructing non-Steiner tree covers for metrics, which is a natural variant of the well-known \EMPH{Steiner point removal} problem for trees (SODA 2001). Given a tree and a set of terminals in the tree,  our goal is to construct a collection of a small number of dominating trees such that for every two points, at least one tree in the collection preserves their distance within a small stretch factor. Here, we identify an unexpected threshold phenomenon around  $2$ where a sharp transition from $n$ trees to $\Theta(\log n)$ trees and then to $O(1)$ trees happens. Specifically, (i) for stretch $ 2-\eps$, one needs $\Omega(n)$ trees; (ii) for stretch $2$, $\Theta(\log n)$ tree is necessary and sufficient; and (iii) for stretch $2+\eps$, a constant number of trees suffice. Furthermore, our lower bound technique for the non-Steiner tree covers of stretch $2$ has further applications in proving lower bounds for two related constructions in tree metrics: reliable spanners and locality-sensitive orderings. Our lower bound for locality-sensitive orderings matches the best upper bound (STOC 2022).

Finally, we study $(1+\eps)$-spanners in planar domains using \EMPH{Steiner points}. In planar domains, Steiner points are necessary to obtain a stretch arbitrarily close to $1$. Here, we construct a $(1+\eps)$-spanner with an \emph{almost linear dependency on $\eps$} in the number of edges; the precise bound is $O((n/\eps)\cdot \log(\eps^{-1}\alpha(n))\cdot \log \eps^{-1})$ edges, where $\alpha(n)$ is the inverse Ackermann function. Our result generalizes to graphs of bounded genus. For $n$ points in a polyhedral metric, we construct a Steiner $(1+\eps)$-spanner with $O((n/\eps)\cdot \log(\eps^{-1}\alpha(n))\cdot \log \eps^{-1})$ edges.
\end{abstract}

\newpage
\setcounter{tocdepth}{2}
	\tableofcontents

\newpage	
\setcounter{page}{1}

\section{Introduction}

Let $\cM = (X,\delta_X)$ be a metric space and $P\subseteq X$ be a set of $n$ points in $\cM$. A \EMPH{$t$-spanner} of $P$ is an edge-weighted graph $G = (P, E, w)$  such that every edge $(p,q)\in E$ has a weight $w(p,q) = \delta_X(p,q)$ and for every two points $x,y\in P$, $\delta_G(x,y)\leq t\cdot \delta_X(x,y)$. Here $\delta_G(x,y)$ denotes the shortest path distance between $x$ and $y$ in $G$. The parameter $t$ is called the \EMPH{stretch} of the spanner $G$. One of the most well-studied class of spanners are Euclidean spanners, where $\cM$ is an Euclidean space. The pioneering work of Chew~\cite{Chew86,Chew89} showed that in the \emph{Euclidean plane} $\mathbb{R}^2$, one can construct a spanner with $O(n)$ edges and $O(1)$ stretch.  Over more than three decades, this result has been refined, improved, and extended in various ways. Most notably, for any $\eps \in (0,1)$, one can construct a spanner with $O(n/\eps)$ edges and stretch $1+\eps$  for point sets in the Euclidean plane~\cite{Clarkson87,Keil88}, and the number of edges is tight~\cite{LS22}. In higher dimensions $d$, one could obtain a similar bound: The number of edges is $O(n/\eps^{d-1})$~\cite{RS91,ADDJS93}, which is also tight~\cite{LS22}.

\begin{figure}[h!]
    \centering
    \includegraphics[width=\textwidth]{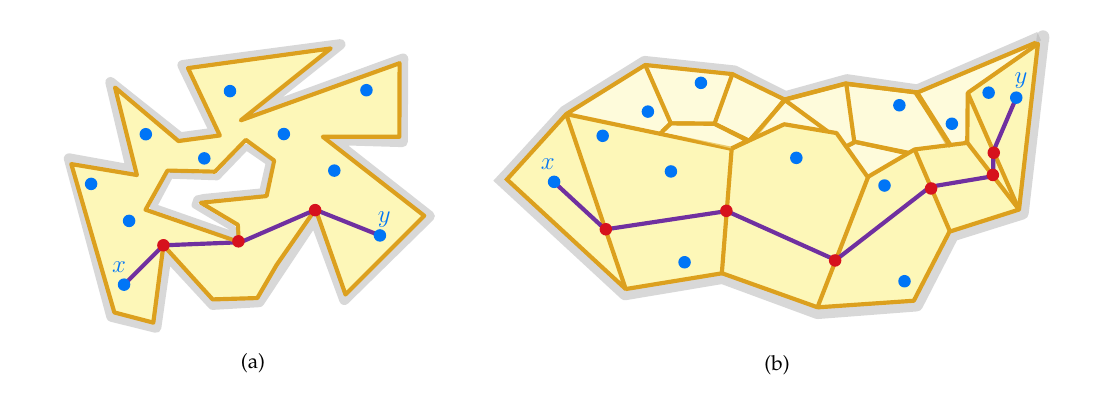}
    \caption{(a) A polygon with holes, blue terminals, and a shortest path between terminals $x$ and $y$. (b) A polyhedral terrain and a shortest path between two points $x$ and $y$.}
    \label{fig:planDom}
\end{figure}

\paragraph{Planar Domains.}  While spanners for points on the Euclidean plane are well understood, in many practical applications, the domain is planar but not Euclidean. One basic example is the \EMPH{polygonal domain} routing in robotics. Here, the metric space $\mathcal{M}$ contains points in a polygon---for example, the floor of a room---and there are (polygonal) obstacles inside the polygon---representing furniture inside the room---called \EMPH{holes}. The distance between two points is measured by the shortest path \emph{avoiding the obstacles}; see \Cref{fig:planDom}(a).  Another important setting is  \EMPH{polyhedral terrain}. A polyhedral terrain is the graph of a piece-wise linear function $f: D\rightarrow \mathbb{R}$  for some convex polygonal region $D\subseteq \mathbb{R}^2$; see \Cref{fig:planDom}(b).  Polyhedral terrains are central in GIS (geographic information system) to model the surfaces of mountains~\cite{HCC11}. Abam, de Berg, and Seraji~\cite{ABS19} noted that polyhedral terrain generalizes polygonal domain. It is relatively easy to show that in both settings, achieving a $(2-\eps)$-spanner for any fixed $\eps\in (0,1)$ requires $\Omega(n^2)$ edges; see Theorem 3 in~\cite{AAHA15}. The main problem is to construct a spanner with stretch $2$ or $2+\eps$ and a linear number of edges.

Abam, Adeli, Homapour, and Asadollahpoor~\cite{AAHA15} constructed a   $(5+\eps)$-spanner for any $n$-point set in a polygonal domain of $h$ holes with $O(n\sqrt{h}\log^2(n))$ edges for any fixed $\eps\in (0,1)$. The number of edges depends on $h$, which could be as large as $n$, and furthermore, there is still a gap in the stretch. Their results were significantly generalized and improved by Abam, de Berg, and Seraji~\cite{ABS19}. They constructed\footnote{There was a technical issue in the proof of Abam, de~Berg, and Seraji~\cite{ABS19}, which was recently fixed by de~Berg, van~Kreveld, and Staals~\cite{DVS23}.} a  $(2+\eps)$-spanner with $O(c(\eps)\cdot n\log n)$  edges, where $c(\eps) = (1+2/\eps)^{O(\log(1/\eps))}$.  Note that the dependence on $\eps$ is \emph{quasi-polynomial}. 

Note that the number of edges of the spanners in both generalized settings~\cite{AAHA15,ABS19} remains $\Omega(n\log n)$ for a constant $\eps>0$, while the number of edges of the spanner in the basic Euclidean setting is $O(n)$. This $\log(n)$ factor gap is due to a fundamental difference in the techniques. The spanner constructions in polygonal domains and polyhedral terrains are based on divide-and-conquer strategy in which $O(n)$ edges will be added in each level of the recursion, resulting in $O(n\log n)$ edges since the recursion depth is $O(\log n)$. On the other hand, in Euclidean spaces, spanner constructions are often non-recursive and directly exploit Euclidean geometry, which is not available in generalized settings. 
 
\begin{question}\label{ques1:polygonal-domain} Can we construct a spanner of stretch $2$ or $2+\eps$ for any fixed $\eps\in (0,1)$ with $O(n)$ edges? Could the dependence on $\eps$, if necessary, be reduced to be polynomial? 
\end{question}

Both positive and negative answers to \Cref{ques1:polygonal-domain} require techniques that are different from those in~\cite{AAHA15,ABS19}. First, for stretch $2$, we show an $\Omega(n\log n)$ lower bound on the number of edges. This is the first super-linear lower bound for stretch $2$. This lower bound suggests that the number of edges for stretch $2+\eps$ could be super linear. Our second positive result shows that this is not the case: we construct a $(2+\eps)$-spanner with $O(n)$ edges for any constant $\eps \in (0,1)$, thus completely answering \Cref{ques1:polygonal-domain}. Our results are summarized in the following theorem.\footnote{The $\Tilde{O}(.)$ notation hides logarithmic factors in $1/\eps$.}

\begin{restatable}{theorem}{TerrainSpanner}\label{thm:spanner-polygonal} 
Let $\eps \in (0,1)$ be a parameter.
\begin{enumerate}
 \item  There exists a polyhedral terrain and a set $P$ of $n$ points on the terrain such that any $2$-spanner for $P$  must have $\Omega(n\log n)$ edges. 
 \item Given any set $P$ of $n$ points in a polyhedral terrain, we can construct a $(2+\eps)$-spanner for $P$ with $\Tilde{O}(n/\eps^6)$  edges.  The number of edges is $O(n)$ for a constant $\eps$.   
\end{enumerate}
\end{restatable}

Our technique for proving \Cref{thm:spanner-polygonal} is by drawing a connection to what we call a non-Steiner tree cover for trees, which will be formally defined in \Cref{sec:techniques}. There, we identify a rather surprising threshold phenomenon around stretch $2$; see \Cref{thm:tree-cover}. The lower bound (item 1) in \Cref{thm:spanner-polygonal} will be given in \Cref{subsec:poly-lb-stretch2} and the upper bound (item 2) construction will be given in \Cref{sec:obstacle}.

\paragraph{Steiner Spanners.}
A complementary direction is to study how Steiner points could help to construct spanners in planar domains. Here, Steiner points are points not in the input point set but in the ambient space. In $\bR^{2}$ (and generally in any $\bR^{d}$), Le and Solomon~\cite{LS22} showed that Steiner points could \emph{quadratically} reduce the dependence of the number of edges of the spanner on $1/\eps$, from $O(n/\eps)$ for non-Steiner spanners to $O(n/\sqrt{\eps})$ for Steiner spanners. Interestingly, the bound $O(n/\sqrt{\eps})$ is tight~\cite{LS22,BT22}. Here, we show that for a polyhedral terrain Steiner points could help in two different ways: They can reduce the stretch from $2+\eps$ to $1+\eps$; and also reduce the dependence on $\eps$. We observe that the construction in~\cite{ABS19} could be used to construct a Steiner $(1+\eps)$-spanner with $O(n\log(n)/\eps^2)$ edges. Using a completely different technique, we almost remove the $\log(n)$ factor and reduce the dependence on $1/\eps$ to linear, which we believe is optimal; see more discussion below.  Furthermore, we completely remove the dependence on $n$ while increasing the dependence on $1/\eps$ to a larger polynomial.

\begin{restatable}{theorem}{SteinerSpanner}\label{thm:steiner} Let $\eps \in (0,1)$ be a parameter. Let $P$ be a set of $n$ points in a polyhedral terrain. We can construct a Steiner $(1+\eps)$-spanner for $P$ with $O\left((n/\eps)\cdot \log(\eps^{-1}\alpha(n))\cdot \log \eps^{-1}\right)$ edges, where $\alpha(n)$ is the inverse Ackermann function. The same result holds even when $P$ is on a polyhedral surface.
\end{restatable}

The proof of \Cref{thm:steiner} will be given in \Cref{sec:lowerbounds}. The high-level idea will be given below.

\subsection{Key Techniques}\label{sec:techniques}

\subsubsection{Non-Steiner Spanners: Proof of \Cref{thm:spanner-polygonal}}

We now describe our technique for constructing $(2+\eps)$-spanners. As alluded to above, breaking the $\Omega(n\log n)$ bound on the number of edges of $(2+\eps)$-spanner requires a novel technique. Our starting point is to understand tree metrics. It is not so hard to see that tree metrics are a special case of polygonal domains.  In this case, we are given a tree $T$ and a subset $P$ of $n$ vertices of $V(T)$, and we want to construct \emph{a sparse structure} on $P$ preserving distances between points in $P$. Here, we introduce a new notion of a sparse structure, called \EMPH{non-Steiner tree cover for tree metrics}; the non-Steiner terminology is used to emphasize that the cover does not use Steiner vertices, those that are in $V(T)\setminus P$.

\paragraph{Non-Steiner Tree Covers for Trees.} In this problem, we are given an edge-weighted tree $T = (V_T,E_T, \omega_T)$, and a set of \EMPH{terminals} $K\subseteq V_T$. We say that a collection $\mathcal{T} = \{T_1,T_2,\ldots ,T_{\beta}\}$ of $\beta$ many edge-weighted trees is an  \EMPH{$(\alpha,\beta)$-non-Steiner tree cover} if the following hold: 
\begin{enumerate}
    \item \textbf{Steiner free.~} For every $i\in [\beta]$, $V(T_i) = K$.
    \item \textbf{Dominating.~} For every $i \in [\beta]$, $d_{T_i}(x,y)\geq d_T(x,y)$ for every two vertices $x,y\in K$.
    \item   \textbf{Low stretch.~} $\min_{i\in [\beta]}d_{T_i}(x,y)\leq \alpha\cdot d_T(x,y)$  for every two vertices $x,y\in K$.
\end{enumerate}
Parameter $\alpha$ is called the \EMPH{stretch} of the cover $\mathcal{T}$, and parameter $\beta$ is called the \EMPH{size} of the cover $\mathcal{T}$.  

Our goal is to construct a non-Steiner tree cover with \emph{small stretch and size}. As we will see later, the size-stretch trade-off for the non-Steiner tree cover of tree metrics is central to our construction of spanners in planar domains. While the weight of an edge $e = (x,y)$ in a tree in the non-Steiner tree cover can be theoretically different from 
$d_T(x,y)$, it is always better to set the weight $w(e)=d_T(x,y)$ since doing so will improve the stretch while preserving all the properties of a non-Steiner tree cover.

The special case of only having exactly one tree in the tree cover is the well-known Steiner Point Removal (SPR) problem for tree metrics introduced by Gupta~\cite{Gupta01}. Gupta achieved stretch at most 8. This result has 
 applications in metric embedding~\cite{CGNRS06,FRT04} and metric labeling~\cite{AFHKTT04}. Gupta also showed a stretch lower bound of $4 -o(1)$. This lower bound was subsequently improved to $8-o(1)$~\cite{CXKR06}, matching the upper bound by Gupta~\cite{Gupta01}.  

Given that stretch 8 is the best possible for one tree, we ask if we could reduce the stretch by using more than one tree. And more generally:

\begin{question}\label{ques1:tree-cover} What is the precise trade-off between the number of trees and the stretch?
\end{question}

Our next result is a complete answer to \Cref{ques1:tree-cover}.

\begin{restatable}{theorem}{TreeCover}\label{thm:tree-cover} Let $\alpha\geq 1$ be the \emph{stretch parameter}, and $\eps\in (0,1)$ be any given constant. Let $T$ be an edge-weighted tree and  $K\subseteq V(T)$ be any set of terminals. 
\begin{enumerate}
    \item If $\alpha = 2+\eps$, then \EMPH{$O(1)$ trees} suffice: we can construct a non-Steiner tree cover for $K$ of size  $O(1)$ and stretch $2+\eps$. The number of trees is $O(\eps^{-2}\log(\eps^{-1}))$.
    \item If $\alpha = 2$, then \EMPH{$O(\log n)$ trees} suffice.  Furthermore, $\Omega(\log n)$ trees are necessary: there exists a tree and a terminal set such that any tree cover with stretch $2$ for the terminals must have $\Omega(\log n)$ trees. 
    \item If $\alpha = 2 - \eps$, then \EMPH{$\Theta(n)$ trees} are both necessary and sufficient. 
\end{enumerate}
\end{restatable}

Here, we also see the threshold phenomenon around stretch $2$: (i) for $\alpha = 2-\eps$, one needs $\Omega(n)$ trees, (ii) for $\alpha = 2$, $\Theta(\log n)$ tree is necessary and sufficient, and (iii) for $\alpha=2+\eps$, a constant number of trees suffice.   We note that both the lower bound and upper bound construction for the third case in \Cref{thm:tree-cover} are very simple. The upper bound is obtained by constructing a single star for each terminal to preserve the distances from the terminal to other terminals. The lower bound is realized by the star graph with terminals being the leaves; this example was also considered by Gupta~\cite{Gupta01}. The upper bound proofs in \Cref{thm:tree-cover} will be given in \Cref{sec:SF-treecover}, and the lower bound of $\Omega(\log n)$ trees will be given in \Cref{subsec:stretch2-treecover}.

Next, we discuss the connection between non-Steiner tree cover and spanners in planar domains.

\paragraph{Spanners in Planar Domains.} 
Let $\bM_1$ and $\bM_2$ be two \emph{families of metric spaces}. We say that \EMPH{$\bM_1\sqsubseteq \bM_2$} if for every metric space $\cM_1\in \bM_1$ (which could be infinite) and any finite point set $P \in \cM_1$, there exists a metric space $\cM_2 \in \bM_2$ such that $P$ embeds isometrically into $\cM_2$. That is, there exists a point set $Q\in \cM_2$ and a bijection $f: P\rightarrow Q$ such that $\delta_1(x,y) = \delta_2(f(x),f(y))$ for every point pair $x,y\in P$. Here $\delta_1$ and $\delta_2$ are the distance functions of $\cM_1$ and $\cM_2$, respectively.  If $\bM_1\sqsubseteq \bM_2$ and $\bM_2\sqsubseteq \bM_1$, we write that \EMPH{$\bM_1 \cong \bM_2$}.  If  $\bM_1\sqsubseteq \bM_2$ and $\bM_2\not\sqsubseteq \bM_1$, then we write  \EMPH{$\bM_1 \sqsubset \bM_2$}.

Let \EMPH{$\trees$}, \EMPH{$\planar$}, \EMPH{$\polydom$}, \EMPH{$\terrain$}, and \EMPH{$\polysurf$} be the family of shortest-path metrics in edge-weighted trees, edge-weighted planar graphs, polygonal domains, polyhedral terrains, and polyhedral surfaces, respectively.  We observe that:

\begin{restatable}{lemma}{MetricRelations}\label{lm:metric-relations} $\trees \sqsubset \planar\cong \polydom \cong \terrain \sqsubset \polysurf$. 
\end{restatable}

Thus, three families of metrics, namely planar metrics, polygonal domains, and polyhedral terrains, are equivalent, and they all strictly contain tree metrics. Abam, de Berg, and Seraji~\cite{ABS19} showed that $\polydom \sqsubseteq \terrain$ by controlling the elevation of polyhedral terrains. We can show that $\terrain\sqsubseteq  \planar$ by looking at the arrangements of the geodesic paths in a polyhedral terrain. Finally, we show that $ \planar\sqsubseteq \polydom$ by using polygonal holes to ``fill in'' the faces of a planar-embedded graph, thereby proving \Cref{lm:metric-relations}; see \Cref{sec:obstacle} for details. Note, however, that planar metrics need not embed in Euclidean spaces (without obstacles); see~\cite{BateniDHM07,NR03,Rao99}.

By \Cref{lm:metric-relations}, we could work with planar metrics instead of polyhedral terrains:  given a set of $n$ terminal points in a planar metric, construct a spanner containing the terminals only. (The planar metric might contain more points than the terminal points.) Specifically, we will use a recent (Steiner) tree cover developed by~\cite{ChangCLMST23}.  A \EMPH{Steiner tree cover}\footnote{Prior work did not include the prefix ``Steiner'' in the Steiner tree cover terminology, since Steiner points are not their focus. Here we clearly distinguish between Steiner and non-Steiner versions of tree covers.} of a metric $\cM = (X,\delta_X)$ is a collection of trees $\mathcal{T}$ such that for every tree $T\in \mathcal{T}$ we have $X \subseteq V(T)$, and $\delta_X(x,y)\leq d_T(x,y)$ for every two points $x,y\in X$. The size of the tree cover $\mathcal{T}$ is the number of trees in $\mathcal{T}$ and the stretch is at most $\alpha$ if $d_T(x,y)) \leq \alpha\cdot \delta_X(x,y)$ for  \emph{some} $T\in \mathcal{T}$.  Note that by definition, $X$ could be a strict subset of $V(T)$, and the points in $V(T)\setminus X$ are called \EMPH{Steiner points}.  In all existing tree cover constructions, Steiner points are \emph{copies} of the points in $X$. A different way to think about this is that a point $p$ in $X$ could appear multiple times in $T$, and we only keep one copy of $p$ in $T$ as the image of $p$, and regard other copies as Steiner points.

\begin{theorem}[Theorem 1.2 in~\cite{ChangCLMST23}]\label{thm:tree-cover-planar} Let $G = (V,E,w)$ be an edge-weighted planar graph and $\eps\in (0,1)$ be any given parameter. We can construct a Steiner tree cover $\mathcal{T}$ for the shortest path metric of $G$ such that (a) $\mathcal{T}$ has stretch $1+\eps$ and (b) $\mathcal{T}$ has $\Tilde{O}(\eps^{-3})$ trees. Furthermore, Steiner points of every tree in $\mathcal{T}$ are copies of points in $X$. 
\end{theorem}

 \Cref{thm:tree-cover-planar} allows us to use non-Steiner tree covers developed in \Cref{thm:tree-cover} to construct a spanner for points in planar metrics, and hence polyhedral terrain by \Cref{lm:metric-relations}. It is worth noting that our spanner is much more structured than simply having a small number of edges: it is the union of a small number of distance-preserving non-Steiner trees. 
 
  Ironically, looking at our series of constructions as a whole, one can see that we first construct a spanner for a point set in a planar domain by \EMPH{adding more} Steiner points (\Cref{thm:tree-cover-planar}) and then \EMPH{removing} all   Steiner points in the final step (\Cref{thm:tree-cover}). 

 We show the lower bound (item 1  of \Cref{thm:spanner-polygonal})  for tree metrics; \Cref{lm:metric-relations} implies that the same lower bound holds for polyhedral terrains. Indeed, our lower bound applies to the comb graphs, with terminals being the leaves.  Here, we establish a connection between non-Steiner spanners for terminals in the comb graphs and low-hop spanners for points in the line metric.  Then, we can slightly adapt the lower bound technique for points in the line metric in~\cite{LMS22} to achieve our result. 

\paragraph{Steiner Spanners: Proof of \Cref{thm:steiner}.}
 Here, we use \Cref{lm:metric-relations} again by constructing a Steiner spanner for a set of terminals in a planar metric. For obtaining a linear number of edges at the cost of a polynomial factor in $1/\eps$, \Cref{thm:tree-cover-planar} suffices: we simply union all the trees in the Steiner tree cover. However, to reduce the dependency on $1/\eps$ to \EMPH{nearly linear} (cf.~\Cref{thm:steiner}), we resort to more advanced tools, including net trees in planar metrics~\cite{LW21}, reduction to additive stretch~\cite{LW21,ChangCLMST23}, and tree shortcutting~\cite{ChungG84,Chazelle87a,AS87,Thorup97,FiltserL22}.  The basic idea is a reduction to constructing a spanner with additive distortion using the net-tree-based spanner technique. We then apply the shortest path separators~\cite{LT79,Thorup04} and tree shortcutting to construct an additive spanner with an almost linear dependency on $1/\eps$.

 We believe that the linear dependency on $1/\eps$ is optimal. For more restricted types of distance-preserving structures, such as minor-free~\cite{KNZ14} or aligned planar structures~\cite{CKT22}\footnote{Refer to page 9 in the arXiv version of~\cite{CKT22} for the exact definition of an aligned planar structure.}, the number of edges was shown to be $\Omega(n/\eps)$ where $n$ is the number of points.

 \subsection{Other Applications}

Here, we present two more applications of our technique for proving lower bound $\Omega(\log n)$ on the number of trees with stretch $2$ in \Cref{thm:tree-cover}. All results claimed here are proven in \Cref{sec:lowerbounds}.

\paragraph{Reliable Spanners.} 
In this problem, we are given a metric space $\cM  = (X,\delta_X)$ and a set $P\subseteq X$ of $n$ points, a $t$-spanner $G$ of $P$ is \EMPH{(deterministic) $\nu$-reliable} for a parameter $\nu\in (0,1)$ if for any subset $B\subseteq  P$, there exists a set $B^+\supseteq B$ of size $|B^+|\leq (1+\nu)\, |B|$ such that $G[P\setminus B]$ is a $t$-spanner of all the points in  $P\setminus B^+$. That is, for every $x,y\in P\setminus B^+$, $\delta_{G[P\setminus B]}(x,y)\leq t\cdot \delta_X(x,y)$. Informally, the spanner is reliable if whenever the vertices in a set ($B$) fail, then it only affects a few other vertices ($B^+\setminus B$).  We say that $G$ is an \EMPH{oblivious $\nu$-reliable} spanner if $G$ is drawn from a distribution $\mathcal{D}$  and $\mathbb{E}_{G\sim \mathcal{D}}[|B^+|] \leq (1+\nu)|B|$. 

Deterministic, reliable spanners were introduced in~\cite{BDMS13} for point sets in Euclidean spaces. Their results were improved greatly by Buchin, Har-Peled, and Ol\'ah~\cite{BuchinHO20}, who constructed a deterministic $\nu$-reliable $(1+\eps)$-spanner of $O\left(n(\log n)(\log\log n)^6\right)$ edges for point sets in $\mathbb{R}^d$ for constants $\eps$, $d$ and $\nu$. This almost matches the lower bound $\Omega(n\log n)$ for $d=1$ by~\cite{BDMS13}. In a follow-up work~\cite{BuchinHO22}, the same authors constructed an oblivious $\nu$-reliable $(1+\eps)$-spanner with $O(n\poly(\log\log(n)))$ edges using \emph{locality sensitive orderings}~\cite{CHJ20}, bypassing the $\Omega(n\log n)$ lower bound for deterministic reliable spanners. 

Har-Peled, Mendel, and Ol\'ah~\cite{HMO23} studied reliable spanners for metric spaces. One basic problem is to construct reliable $t$-spanners for tree metrics. Specifically, for constants $\nu$ and $\eps$,  they constructed oblivious $\nu$-reliable spanners with stretch $3+\eps$ and $ O(n \log^2n \log^2\Phi \log(\log(\Phi)\log(n)))$  edges where $\Phi$ is the aspect ratio of the metric. Note that there exists a tree metric such that any oblivious reliable spanner with stretch $2-\eps$ for any constant $\eps \in (0,1)$ must have $\Omega(n^2)$ edges~\cite{FL22B}. Filtser and Le~\cite{FL22B} reduced the stretch to $2$ (which is optimal) and the number of edges to $O\left(n\log^5(n)\right)$  by developing a variant of locality-sensitive orderings for tree metrics. The main open problem is how many edges are necessary and sufficient for stretch $2$. It is conceivable that, similar to the Euclidean case, $O(n\poly(\log\log n))$ edges suffice for tree metrics, given that the techniques in these settings are quite similar. Here, we show this is not the case by proving an $\Omega(n\log n)$ lower bound for the number of edges using the technique developed for \Cref{thm:tree-cover}. Our result separates Euclidean metrics from tree metrics.

\begin{restatable}{theorem}{ReliableSP}\label{thm:reliable-lb} 
There exists a tree metric $T$ with $n$ points such that any oblivious $\frac{1}{3}$-reliable $2$-spanner for $V(T)$ must have $\Omega(n\log n)$ edges. 
\end{restatable}

\paragraph{Locality Sensitive Ordering.} 
Chan, Har-Peled, and Jones~\cite{CHJ20} introduced the notion of \EMPH{locality-sensitive ordering} (\EMPH{LSO}) and showed that for any point set in $\mathbb{R}^d$, one can construct a locality-sensitive ordering, comprised of $O(1)$ orderings for constant dimension $d$ and stretch parameter $\eps \in (0,1)$. Their LSO has many surprising algorithmic applications, including dynamic spanners, dynamic approximate minimum spanning trees, dynamic bichromatic closest pairs, approximate nearest neighbors~\cite{CHJ20}, and reliable spanners~\cite{BuchinHO22}.  However, for tree metrics, their notion of LSO is too strong: one needs $\Omega(n)$ orderings in an LSO for any fixed $\eps \in (0,\frac12)$. Filtser and Le~\cite{FL22B} introduced a more relaxed version of LSO tailored specifically for tree metrics, called \emph{left-sided LSO}. As defined in~\cite{FL22B}, a \emph{$(\tau,\rho)$-left-sided LSO} for a tree metric $T$ of $n$ points is a collection $\Sigma$ of  \emph{linear orderings} over \emph{subsets} of $V(T)$ such that (i) every point $x$ in $T$ belongs to at most $\tau$ linear orderings, and (ii) for any two points $x,y \in T$, there exists an ordering $\sigma \in \Sigma$ with the following property: for any $x'\preceq_{\sigma} x$ and $y'\preceq_{\sigma} y$, $\delta_T(x',y') \leq  \rho \delta_T(x,y)$. That is,  the distance of any two points \emph{to the left} of $x$ and $y$ in $\sigma$ is at most $\rho\cdot\delta_T(x,y)$. Parameter $\tau$ is the \EMPH{size} of the ordering $\Sigma$ and $\rho$ is the \EMPH{stretch}. Filtser and Le~\cite{FL22B} showed that tree metrics admit a left-sided LSO with $O(\log n)$ size and stretch $\rho = 1$. A question raised by their work is: Could the size of the ordering be reduced to a $O(1)$? We answer this question negatively by showing that $O(\log n)$ size is indeed \emph{optimal}. We do so by using the technique developed in the proof of \Cref{thm:tree-cover}.

\begin{restatable}{theorem}{LeftLSO}\label{thm:left-LSo} There exists a tree metric $T$ with $n$ points such that any $(\tau,\rho)$-left-sided LSO for $T$ with $\rho = 1$ must have $\tau = \Omega(\log n)$, matching the $O(\log n)$ upper bound by Filtser and Le~\cite{FL22B}.
\end{restatable}

\subsection{Further Related Work}

Steiner spanners were studied for point sets in Euclidean plane with obstacles~\cite{ACCDSZ}, which is the same as polygonal domains, in the context of querying obstacle-avoiding shortest paths. In this setting, the \emph{vertices of the polygonal obstacles belong to the point set}. Arikati et al.~\cite{ACCDSZ} constructed a \emph{planar} Steiner $t$-spanner for $L_1$ distance with stretch $t = 1+\eps$ and $O(n/\eps^2)$ edges; and for $L_p$ distance with stretch $t = 2^{(p-1)/p} + \eps$ and $O_{\eps}(n)$ edges where the $O_{\eps}(.)$ notation hides the dependence on $\eps$, which was not explicitly computed in~\cite{ACCDSZ}. Specifically for Euclidean distance, the stretch is $\sqrt{2}+\eps$. Our spanner in \Cref{thm:steiner} has stretch $1+\eps$ and almost linear dependence on $1/\eps$; furthermore, the number of edges of our spanner \emph{does not} depend on the number of vertices of the obstacles, which may be arbitrary.

Kapoor and Li~\cite{KL09} constructed a Steiner $(1+\eps)$-spanner for points in a polyhedral surface $\mathcal{P}$, which is a higher-genus generalization of polyhedral terrains (cf.~\Cref{lm:metric-relations}). Their spanner has $O(\gamma(\mathcal{P})n/\eps)$ edges, where $\gamma(\mathcal{P})$ is the \emph{geodesic dilation factor} of the surface $\mathcal{P}$, which measures how nice $\mathcal{P}$ is. In the worst case, $\gamma(\mathcal{P})$ could be up to $\Theta(n)$, and hence the spanner has a trivial $O(n^2)$ edges.  Our spanner in \Cref{thm:steiner} has linearly many edges regardless of $\gamma(\mathcal{P})$. 

Another related direction is to study the  \emph{complexity} of geodesic spanners in planar and polyhedral domains. An edge in our spanner might be realized by a geodesic path of up to $\Omega(n)$) edges in the input domain. For example, in a polygonal domain, an edge $(u,v)$ in our spanner could only be realized by a geodesic obstacle-avoiding path of many straight-line edges in the domain. The total number of edges in the input domain to ``realize'' our spanners is called the \EMPH{complexity} of the spanners. This question has recently been studied in depth for both non-Steiner spanners \cite{DVS23} and the Steiner version~\cite{dOPSW24}. The main finding is that, for any $k\geq 1$, a $(2k+\eps)$-spanner for $n$ points in a simple polygon with $m$ vertices has complexity $O(mn^{1/k} +n \log^2n)$~\cite{DVS23}, and that using Steiner points does not help to reduce the complexity by much~\cite{dOPSW24}.  

\section{Non-Steiner Tree Cover for Trees}\label{sec:SF-treecover}

In this section, we prove \Cref{thm:tree-cover}, which we restate below.

\TreeCover*

By scaling, we may assume that the \EMPH{minimum distance between any two vertices of $T$ is at least $8$} and the maximum distance between any vertices is at most $\Delta$ for an integer $\Delta$.  (The distance lower bound of $8$ is somewhat arbitrary; any sufficiently large constant works.) 

We view the edge-weighted tree $T = (V_T,E_T, w_T)$ as a \emph{continuous tree} by viewing each edge $(u,v)$ as a continuous line segment (i.e., a geometric realization of a 1D cell complex). We still use \emph{vertices} to refer to the vertices of the discrete $T$. The distances between points corresponding to vertices of $T$ are the distances in the tree, and the distance between any two points in the same line segment of an edge is the length of the sub-segment of the edge connecting the two points. We could naturally extend the distance function to measure the distance between \emph{any} two points $p_1,p_2\in T$ as follows: let $(u_i,v_i)$ be the edge containing $p_i$  for $i = 1,2$; we assign
\begin{equation}
    d_T(p_1,p_2) = \min  \begin{cases}
    d_T(p_1,u_1) + d_T(u_1,u_2) +    d_T(u_2,p_2),    \\
    d_T(p_1,u_1) + d_T(u_1,v_2) +    d_T(v_2,p_2),  \\
    d_T(p_1,v_1) + d_T(v_1,v_2) +    d_T(v_2,p_2),  \\
    d_T(p_1,v_1) + d_T(v_1,u_2) +    d_T(u_2,p_2)  .
  \end{cases}
\end{equation}

In a single tree construction by Gupta~\cite{Gupta01}, one could assume that the terminals are in the leaves of $T$. However, here we could not make the same assumption when constructing more than one tree in the cover, making our construction more complicated. 

To illustrate our new ideas, we first present the construction of a non-Steiner tree cover of stretch $3+\eps$ in \Cref{sec:dist-3-cover}; the tree cover has size $O(1/\eps\log(1/\eps))$. Then, in \Cref{sec:dist-2e-cover}, we show how to reduce the stretch to $2+\eps$ at the expense of another factor of $1/\eps$ in the number of trees using additional insights. In \Cref{sec:dist-2-cover}, we construct a tree cover with $O(\log n)$ trees and stretch exactly $2$.   

\subsection{Stretch $3+\eps$}\label{sec:dist-3-cover}

We will show a construction with stretch $3+12\eps$; we could recover stretch $3+\eps$ by scaling $\eps$. Suppose that we root $T$ at a non-terminal point $r_0$. Our construction is based on what we call \EMPH{$\eps$-chops}. We assume that $\eps \in (0,1)$. For every point $x\in T$, let $C(x)$ be the terminal closest to $x$ in the tree $T$, breaking ties \emph{consistently}, that is, according to some universal linear ordering on the terminals. The closest terminal might not be in the subtree rooted at $x$. (This is in contrast to Gupta's construction~\cite{Gupta01} where $C(x)$ was defined to be the closest terminal in the subtree rooted at $x$.) Let $h(x) = d_T(x,C(x))$, which is the distance from $x$ to its closest terminal. We choose two parameters:

\begin{equation}\label{eq:k-p-def}
     k \text{ the smallest integer such that } (1-\eps)^{k} \leq \eps, \qquad \text{and  } p = 1/\eps+1.
\end{equation}
Note that $k = \Theta(1/\eps\log(1/\eps))$. Let:
\begin{equation}\label{eq:def-hep}
    h_\eps(x) = \begin{cases} 
   \eps \cdot h(x) & \text{if } h(x) \geq (1-\eps)^{k\cdot p}  \\
   \eps      & \text{otherwise.}
  \end{cases}
\end{equation}

We note that $(1-\eps)^{k\cdot p} \approx \eps^{\Theta(1/\eps)} $ for the choice of $k$ and $p$ in \Cref{eq:k-p-def}.   Roughly speaking, $h_\eps(x)$ is about $\eps h(x)$ unless when $h(x)$ is smaller than $\eps^{\Theta(1/\eps)}$. 

\begin{figure}[h!]
    \centering
    \includegraphics[width=\textwidth]{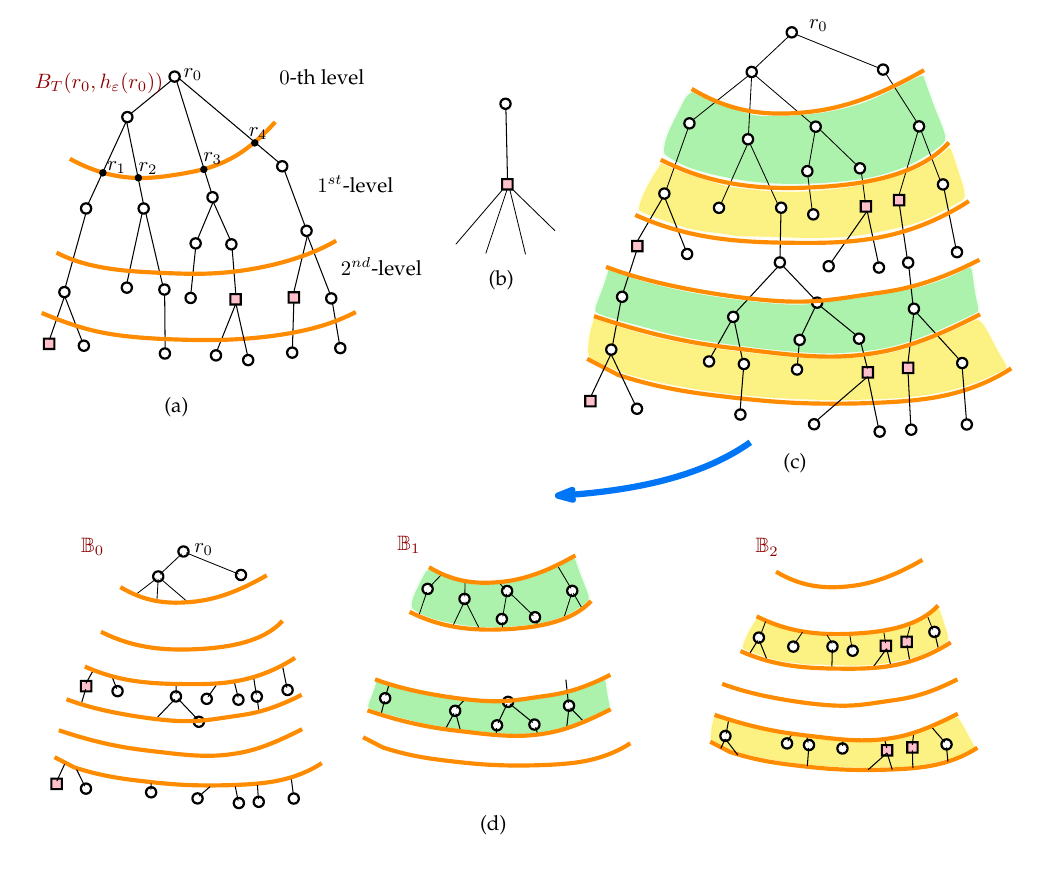}
    \caption{(a) $\eps$-chops; (b) a jump chop; (c) a collection of chops $\mathbb{C}$ for $k = 3$ that is partitioned into 3 buckets $\bB_0,\bB_1,\bB_2$ in (d). Square vertices are terminals. }
    \label{fig:chops}
\end{figure}

\begin{quote}
    \textbf{$\eps$-Chops.~} Let $B_T(r_0, h_{\eps}(r_0))$ be the ball of radius $h_{\eps}(r_0)$ centered at $r_0$, which contains all points in $T$ within distance at most $h_{\eps}(r_0)$ from $r_0$. We then ``chop'' the tree $T$ by removing $B_T(r_0, h_{\eps}(r_0))$ from $T$. We call  $B_T(r_0, h_{\eps}(r_0))$ a 0-th level \EMPH{$\eps$-chop} rooted at $r_0$. Note that $B_T(r_0,h_{\eps}(r_0))$ induces a connected subtree of $T$. Let $T_1,\ldots, T_a$ be the resulting subtrees of $T$ after removing  $B_T(r_0, h_{\eps}(r_0))$, with roots $r_1\ldots, r_a$, respectively (see \Cref{fig:chops}(a)). We then recursively chop each tree  $T_i$ by removing a ball of radius $h_{\eps}(r_i)$, which is $B_{T_i}(r_i, h_{\eps}(r_i))$, from the root $r_i$ for every $i \in [a]$. Each tree $B_{T_i}(r_i,h_{\eps}(r_i))$ is called a 1st level \EMPH{$\eps$-chop} rooted at $r_i$. We repeat the process for each remaining subtree until every point of $T$ is chopped at some level. We denote the result of the chopping process by $\mathbb{C} = \{\cL_0,\cL_1, \ldots\}$ where $\cL_i$ contains $\eps$-chops of $T$ at level $i$. 
\end{quote}

Ideally, we want every chop at a root $r$ to have radius $\eps h(r)$. However, if we do so, we will never chop a terminal; as the chopping process gets closer to the terminal, the radius of the chop gets smaller. So we fix this by imposing a chop of radius $\eps$ whenever $h(r)$ becomes smaller than $(1-\eps)^{pk}$. If $h_\eps(r)  = \eps h(r)$, we call the chop a \EMPH{regular chop}; otherwise,  $h_\eps(r)  = \eps$, and we call the chop a \EMPH{jump chop}; see \Cref{fig:chops}(b).   The jump chop is a technical fix that we introduce to handle the case where internal nodes of the tree could be terminals, as alluded to above. 

We could define an ancestor-descendant relationship between two $\eps$-chops: We say that an $\eps$-chop $X$ is an \EMPH{ancestor} of an $\eps$-chop $Y$ if the root of $X$ is an ancestor of the root of $Y$. If $Y\not= X$, we say that $X$ is a \EMPH{proper ancestor} of $Y$.  

\begin{observation}\label{obs:chop}  Let $\mathcal{L}_i$ and $\mathcal{L}_j$ be two chops at level $i < j$. Then, for $\eps$-chop $X\in \mathcal{L}_j$, there exists a unique $\eps$-chop $Y\in \mathcal{L}_i$ such that $X$ is a descendant of $Y$. 
\end{observation}

\begin{claim}\label{clm:jump-chop} Let $X$ be an $\eps$-chop rooted at $x$. Then $X$ contains at most one terminal, and $X$ contains a terminal if and only if it is a jump chop. Furthermore, the terminal in $X$, if any, is the closest terminal of $x$ in $T$ and is the vertex (of the discrete tree $T$) in $X$.
\end{claim}
\begin{proof}
    If $X$ is a regular chop rooted at $x$ then then the radius of the chop is $\epsilon h(x)<h(x)\le d(x,t)$ for any terminal $t$, and thus $t$ cannot belong to $X$. If $X$ is a jump chop and it contains a terminal $t$, then, for any other terminal $t_0$ in $X$, we have $d(t_0,x)\ge d(t,t_0)-d(x,t_0)\ge 8-\epsilon$ by the triangle inequality, the definition of the jump chop and the fact that distances between vertices in the tree are at least $8$. Thus, clearly, $t_0$ cannot belong to $X$ and also $t$ is the closest terminal to $x$.  
\end{proof}

\begin{figure}[h!]
    \centering
    \includegraphics[width=\textwidth]{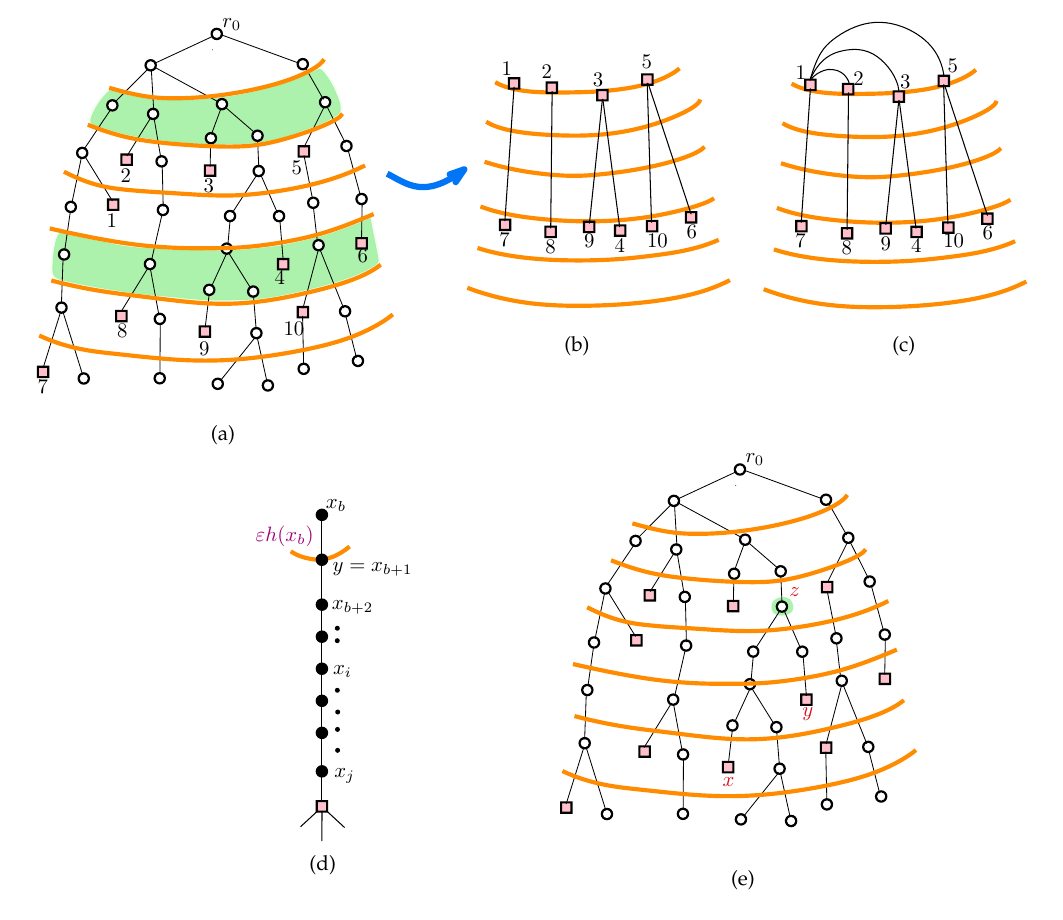}
    \caption{(a) The second bucket $\bB_2$; (b) and (c) a tree constructed from $\bB_2$; (d) all the $\eps$-chops close enough to a jump chop are segments on the same edge; and (e) $z = \lca(x,y)$ belongs to a chop at some level $i+sk$.}
    \label{fig:treecover}
\end{figure}

\paragraph{Cover construction.~} We partition the chops in $\mathbb{C}$ into $k$ buckets $\bB_0, \ldots \bB_{k-1}$ where $\bB_i$ contains chops at level $i$ modulo $k$; see \Cref{fig:chops}(c). More precisely:
\begin{equation}\label{eq:bucket3-def}
    \bB_i = \{\cL_j: j\equiv i \mod k\}.
\end{equation}

For each bucket $\bB_i$, we construct a tree $T_i$ as follows: 

\begin{quote}
    \textbf{Constructing $T_i$.} We consider chops in $\bB_i$ from lower levels to higher levels.  For each $\eps$-chop $X$ in $\cL_{i+sk}$ (at level $i+sk$), where $s$ is an integer, we do the following. First, let $s\geq 1$. Let $x$ be the root of $X$.  Let $Y$ be chop in $\cL_{i+(s-1)k}$ that is the ancestor of $X$; $Y$ exists by \Cref{obs:chop}.   We then add terminal $C(x)$ to $T_i$ and connect $C(x)$ to $C(y)$ with an edge. (The terminal $C(y)$ was added when the algorithm considered $Y$ in the previous step.) Next, if $s = 0$, then we simply add a terminal $C(x)$ corresponding to the root of every tree $X$ in the chop at level $i$.  At this point, $T_i$ is a forest---see \Cref{fig:treecover}(a) and (b)---where every tree is rooted at the terminal corresponding to trees at level-$i$ chop. Finally, we could designate an (arbitrary) root $t$ of a tree in $T_i$ and make the roots of other trees children of $t$; see \Cref{fig:treecover}(c).  Now, $T_i$ is a tree that contains only terminals. We then set of the weight of every edge $(t_1,t_2)$ in $T_i$ to be $d_T(t_1,t_2)$.
\end{quote}

Our tree cover is $\mathcal{T} = \{T_0,T_1,\ldots, T_{k-1}\}$. We note that in this construction, one terminal could appear multiple times in $T_i$, but the copies of the same terminal will form a subtree of $T_i$ (with edges of weight 0) since we break ties consistently. Thus, we could contract all of them into a single terminal. Here, we keep the copies separate to simplify the stretch analysis. 

At this point, it might not be clear why each tree $T_i$ contains all terminals since a terminal $t$ might be chopped by some chop in another bucket $\bB_j$ for $j\not= i$ and hence might not be present in any tree in the chops in $\bB_i$. However, we will show below that in this case, $t$ will be used to replace the root of some tree in $\bB_i$, and hence will be present in $T_i$.

First, we observe that, as we approach a terminal $t$ in the chopping process, the roots of $\eps$-chops close to $t$ will all be replaced by $t$. 

\begin{lemma}\label{lm:terminal-consecutive} Let $t$ be any terminal in $T$. Let $X_i,X_{i+1}, \ldots, X_{j}$ be a sequence of $\eps$-chops at \emph{consecutive levels} rooted at $x_i,x_{i+1},\ldots, x_j$, respectively, such that: (i) $X_j$ contains $t$, (ii) $X_a$ is an ancestor of $X_{a+1}$ for any $i\leq a \leq j-1$, and (ii) $j - i\leq pk-1$. Then $X_i,X_{i+1}, \ldots, X_{j}$ are segments of the same edge (in the discrete $T$), and  $C(x_i) = C(x_{i+1}) = \ldots = C(x_j) = t$.
\end{lemma}
\begin{proof}
    Let us first note that, for any terminal $t_0$ and $s\in[i,j-1]$ such that $X_s$ is a regular chop, by the triangle inequality we have $$d_T(x_{s+1},t_0)\ge d_T(x_{s},t_0)-\epsilon h(x_s)\ge (1-\eps) d_T(x_{s},t_0).$$

    Since $X_j$ contains $t$, then by Claim~\ref{clm:jump-chop}, $X_j$ must be a jump chop and $C(x_j) = t.$ Therefore, by the definition of $h_{\eps}(x_j),$ we have $d_T(x_j,t) = d_T(x_j, C(x_j)) <(1-\eps)^{kp}$. Assume that there is another jump chop among $X_i,\ldots, X_j$, and, moreover, $\ell\in [i,j-1]$ is the largest index of the jump chop. Then $X_{\ell+1},\ldots, X_{j-1}$ are regular chops. Using the displayed inequality above for $s=j-1,j-2,\ldots, \ell+1$ with $t_0 := t$, we get that $$d_T(x_{\ell+1},t)\le (1-\eps)^{\ell+1-j}(1-\eps)^{kp}\le 1-\epsilon.$$ Here we used that $j-\ell-1\le j-i\le pk-1.$ Thus, by the triangle inequality, $d_T(x_{\ell},t)\le d_T(x_{\ell+1},t)+\eps\le 1.$ From here, we see that, for any terminal $t_0\ne t$ we have $d_T(x_{\ell},t_0)\ge d_T(t,t_0)-d_T(x_{\ell},t) \ge 8-1 = 7$ by the triangle inequality. That is, $C(x_{\ell}) =t$. On the other hand, $t$ is contained in $X_j$ and thus $X_\ell$ cannot contain $t$, implying that $d_T(x_{\ell},t)>\eps$. Thus, $x_\ell$ is too far from a terminal, which is a contradiction with the definition of a jump chop.
    
    We conclude that all chops $X_i,\ldots, X_{j-1}$ are regular chops. By the same analysis as above, we have $d_T(x_{s},t)\le 1$  and $C(x_s) = t$ for each $s\in [i,j]$. Finally, the vertices $x_s$, $s\in [i,j-1]$ must lie  on the path from $t$ to the root $r_0$, and, given that $d_T(x_i,t)\le 1,$ must lie on the edge $(u,t)$, where $u$ is the parent of $t$. \end{proof}

A direct corollary is that every terminal will appear in some tree $T_i$. 

\begin{corollary}\label{cor:terminal-cov} Let $t$ be any terminal in $T$, and $\bB_i$ be a bucket for some $i\in \{0,1\ldots, k\}$.  Let $s\geq 0$ be the largest integer such that there exists an $\eps$-chop $X_{i+sk}$ at level $i+sk$ containing an ancestor of $t$, meaning that there exists a point in $X_{i+sk}$ on the path from $t$ to the root of $T$. Then $C(x_{i+sk}) = t$ where $x_{i+sk}$ is the root of $X_{i+sk}$.
\end{corollary}
\begin{proof}
     Let $X_{i+sk}, X_{i+1+sk},\ldots X_{j+sk}$ be the $\eps$-chops at consecutive levels such that $X_{j+sk}$ is the chop containing $t$, and one chop is the ancestor of the next in the sequence.  Let $x_{a+sk}$ be the root of $X_{a+sk}$ for every $i\leq a \leq j$. By the choice of $s$, $|j - i|\leq k-1$ and hence $|j-i|\leq pk-1$ as $p\geq 1$.  By \Cref{lm:terminal-consecutive}, $C(x_{i+sk}) = C(x_{i+1+sk}) = \ldots = C(x_{j+sk})= t$, as claimed.
\end{proof}

\paragraph{Stretch analysis.~} Let $x,y$ be any two terminals  and $z = \lca(x,y)$ be the lowest common ancestor of $x,y$ in $T$. There must be a chop $\cL_{i+{sk}}$ at level $i+sk$ for some $i\in \{0,\ldots k-1\}$ and $s\geq 0$ such that $z$ belongs to the chop. Let $r_{i+sk}$ be the root of the subtree in $\cL_{i+{sk}}$ containing $z$. The next claim allows us to focus our attention to analyzing the distance from $x$ to $r_{i+sk}$ (and symmetrically, the distance from $y$ to $r_{i+sk}$). See \Cref{fig:treecover}(e).

\begin{claim}\label{clm:xy-vs-xroot} $d_T(x,y)\geq (1-\eps)d_T(x,r_{i+sk}) + (1-\eps)d_T(y,r_{i+sk}) - 2\eps$.
\end{claim}
\begin{proof} Observe that
\begin{equation*}
    \begin{split}
        d_T(x,y)& = d_T(x,z) + d_T(z,y)\\
        &\geq  d_T(x,r_{i+sk}) + d_T(r_{i+sk},y) - 2d_T(r_{i+sk},z) \\
        &\geq  d_T(x,r_{i+sk}) + d_T(r_{i+sk},y) - 2h_{\eps}(r_{i+{sk}}). 
    \end{split}
\end{equation*}
By definition, $h_{\eps}(r_{i+{sk}}) \leq \eps + \eps\cdot h(r_{i+sk})$ and furthermore, $h(r_{i+sk}) = d_T(r_{i+sk}, C(r_{i+sk}))\leq d_T(r_{i+sk}, x)$ since $x$ is a terminal. By the same argument, $h(r_{i+sk}) \leq d_T(r_{i+sk}, y)$. Combining with the above equation, we have
\begin{equation*}
    \begin{split}
        d_T(x,y) & \geq d_T(x,r_{i+sk}) + d_T(r_{i+sk},y)  - (\eps + \eps d_T(r_{i+sk}, x)) -   (\eps + \eps d_T(r_{i+sk}, y))\\
        &\geq (1-\eps)d_T(r_{i+sk}, x) + (1-\eps) d_T(r_{i+sk}, y) - 2\eps,
    \end{split}
\end{equation*}
as claimed.
\end{proof}

We remark that $d_T(x,y)\geq 8$ and so the $2\eps$ term in \Cref{clm:xy-vs-xroot} is at most $\eps d_T(x,y)$.

\begin{figure}[h!]
    \centering
    \includegraphics[width=\textwidth]{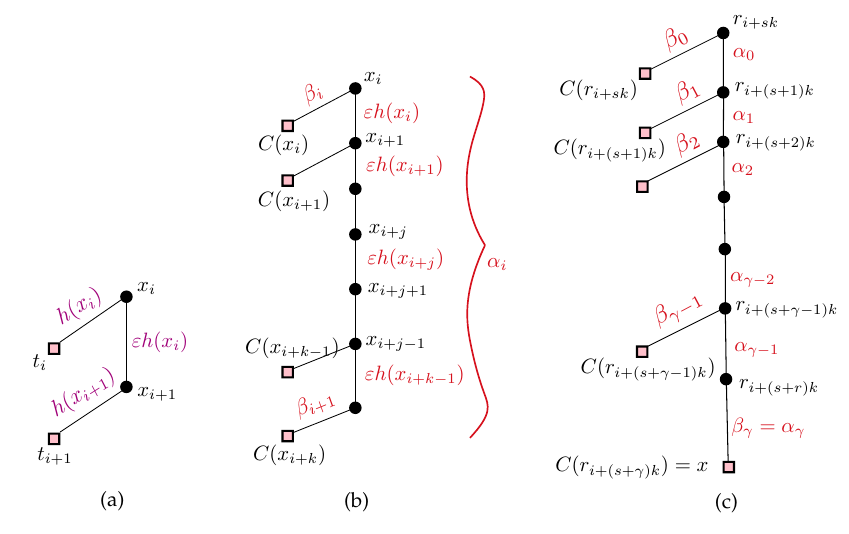}
    \caption{(a) illustration for \Cref{clm:reg-chop-decrease};  (b) illustration for \Cref{lm:dist-chops-modk}; (c) illustration for \Cref{lm:dist-x-r_isk}. }
    \label{fig:dist3}
\end{figure}

We have the following property of the regular chops. 

\begin{claim}\label{clm:reg-chop-decrease} Let $X_i$ be a regular $\eps$-chop rooted at $x$ at some level-$i$. Let $X_{i+1}$ be a descendant chop of $X_i$ rooted $x_{i+1}$ at level $i+1$. Then $h(x_{i+1})\geq (1-\eps)h(x_i)$ and $h(x_{i+1})\leq (1+\eps)h(x_i)$. 
\end{claim}
\begin{proof} Let $t_{i}$ and $t_{i+1}$ be the closest terminal to $x_i$ and $x_{i+1}$, respectively; see \Cref{fig:dist3}(a). Note that $d_T(x_i,x_{i+1}) = \eps h(x_i)$ since $X_i$ is a regular chop.  Then  by definition of $t_{i}$, we have:
\begin{equation*}
\begin{split}
   h(x_i) &= d_T(x_i,t_i) \leq d_T(x_{i}, t_{i+1}) \\
    &\leq    d_T(x_{i}, x_{i+1}) + d_T(x_{i+1}, t_{i+1}) \\
&= \eps h(x_i) +  d_T(x_{i+1}, t_{i+1}) \\
&= \eps h(x_i) + h(x_{i+1})~.
\end{split}
\end{equation*}

For the second inequality, we have:
\begin{equation*}
    \begin{split}
        h(x_{i+1}) &\leq d_T(x_{i+1}, t_i) \leq d_T(x_{i+1}, x_{i}) + d_T(x_{i}, t_{i}) =  (1+\eps)h(x_i),
    \end{split}
\end{equation*}
as claimed.
\end{proof}

\begin{lemma}\label{lm:dist-chops-modk} Let $X_i$ be a chop at some level-$i$ chop $\cL_i$ with root $x_i$. Let $X_{i+k}$ be a descendant chop of $X_i$ at level-$(i+k)$ chop $\cL_{i+k}$ with root $x_{i+k}$. Let $\beta_i = d_T(x_i, C(x_i))$ and $\alpha_i = d_T(x_i, x_{i+k})$ and $\beta_{i+1} = d_T(x_{i+k}, C(x_{i+k}))$. Then:
\begin{enumerate}
    \item $\beta_i \leq (1+2\eps)\alpha_i$ when $\eps\leq 1/2$.
    \item  $\beta_{i+1}\leq \alpha_{i} + \beta_i$.
\end{enumerate}

\end{lemma}
\begin{proof} See \Cref{fig:dist3}(b). We observe that the second item follows from the triangle inequality:  
\begin{equation*}
    \alpha_{i} + \beta_i \geq d_T(x_{i+k}, C(x_i))\geq d_T(x_{i+k}, C(x_{i+k})) = \beta_{i+1}~.
\end{equation*}
We focus on proving the first item. Let $X_{i+j}$ for $1\leq j \leq k-1$ be the descendant chops of $X_i$ and ancestor chops of $X_{i+k}$; $X_{i+j}$ is at level $i+j$. Let $x_{i+j}$ be the root of $X_{i+j}$. If any of the chops between $X_{i}$ and $X_{i+k}$, excluding $X_{i+k}$, say $X_{i+b}$ for some $b\in [0,k-1]$, is a jump chop, which contains a terminal $t$, then $C(x_{i+j}) = t$ for every $i\in [0,b]$ by \Cref{lm:terminal-consecutive}. And furthermore the path from $x_{i}$ to $x_{i+k}$ will go through $t$ as $X_{i+b}$ is the ancestor of $X_{i+k}$. This means $\beta_i\leq \alpha_i$, and item 1 holds.

    Thus, we only consider the case where every chop between  $X_{i}$ and $X_{i+k}$, excluding $X_{i+k}$, is regular. Thus, each $X_{i+j}$ is formed by chopping the subtree rooted at $x_{i+j}$ with radius $\eps h(x_{i+j})$ for every $j \in [0,k-1]$. Thus, we have $\alpha_i = \eps \sum_{j=0}^{k-1} h(x_{i+j})$. By \Cref{clm:reg-chop-decrease}, $h(x_{i+j+1})\geq (1-\eps)h(x_{i+j})$. Thus, we get
    \begin{equation*}
    \begin{split}
      \alpha_i  &\geq \eps (h(x_i) + (1-\eps) h(x_{i}) + \ldots + (1-\eps)^{k-1} h(x_{i}))    \\
      &\geq \eps \beta_i(1+(1-\eps) + \ldots + (1-\eps)^{k-1})\\
      &= \beta_i(1 - (1-\eps)^{k})\geq \beta_i(1-\eps) .
    \end{split}
    \end{equation*}
by the choice of $k$ in \Cref{eq:k-p-def}. This gives $\beta_i\leq \alpha_i/(1-\eps) \leq (1+2\eps)\alpha_i$ when $\eps\leq \frac12$.
\end{proof}

With \Cref{lm:dist-chops-modk} at hand, we are ready to bound $d_{T_i}(r_{i+sk}, x)$. 
\begin{lemma}\label{lm:dist-x-r_isk} $d_{T_i}(C(r_{i+sk}), x)\leq (3+4\eps)d_T(r_{i+sk}, x)$. Similarly, $d_{T_i}(C(r_{i+sk}), y)\leq (3+4\eps)d_T(r_{i+sk}, y)$.
\end{lemma}
\begin{proof}
Let $\{r_{i + (s+1)k}, r_{i + (s+2)k}, \ldots, r_{i+(s+\gamma)k}\}$ be the set of all the roots of the chops in $\bB_i$ that are on the path from $r_{i + sk}$ to $x$. See \Cref{fig:dist3}(c). Note that if $\gamma = 0$, then by \Cref{cor:terminal-cov}, $C(r_{i+sk}) = x$ and hence $d_{T_i}(C(r_{i+sk}), x) = 0$, so the lemma trivially holds. We now assume that $\gamma\geq 1$.

By \Cref{cor:terminal-cov}, $C(r_{i + (s+\gamma)k}) = x$.  By construction, we have
\begin{equation}\label{eq:dist-x-r_isk}
\begin{split}
    d_{T_i}(x,C(r_{i+sk})) &= d_{T_i}(C(r_{i + (s+1)k}), C(r_{i + sk})) +  \ldots + d_{T_i}(C(r_{i + (s+\gamma)k}), C(r_{i + (s+\gamma-1)k})) \\
    &=  d_{T}(C(r_{i + (s+1)k}), C(r_{i + sk}))  + \ldots + d_{T}(C(r_{i + (s+\gamma)k}), C(r_{i + (s+\gamma-1)k})) .
\end{split}
\end{equation}

Let $\beta_a = d_T(r_{i+(s+a)k}, C(r_{i+(s+a)k}))$ for every $0\leq a\leq \gamma$ and $\alpha_a = d_T(r_{i+(s+a)k}, r_{i+(s+a+1)k})$ for every $0\leq a \leq \gamma-1$. Let $\alpha_\gamma =d_T(r_{i+(s+\gamma)k},x)$; by \Cref{cor:terminal-cov}, $\alpha_\gamma = \beta_\gamma$. By \Cref{lm:dist-chops-modk} and the fact that $\alpha_\gamma = \beta_\gamma$, we have
\begin{equation}\label{eq:alpha-vs-beta}
    \beta_a \leq (1+2\eps)\alpha_a\quad \forall ~ 0\leq a \leq \gamma .
\end{equation}

Combing \Cref{eq:dist-x-r_isk} and the triangle inequality, we obtain:

\begin{equation}\label{eq:stretch-3}
    \begin{split}
         d_{T_i}(x,C(r_{i+sk})) &\leq   (\beta_0 + \alpha_0 + \beta_1) + (\beta_1 + \alpha_1 + \beta_2)  + \ldots + (\beta_{\gamma-1} + \alpha_{\gamma-1} + \beta_\gamma)  \\
         &=  \sum_{a=0}^{\gamma-1}\alpha_a +  2\sum_{a = 1}^{\gamma-1}\beta_a + \beta_\gamma + \beta_0\\
         &\leq \sum_{a=0}^{\gamma}\alpha_a  + 2\sum_{a = 0}^{\gamma}\beta_a \\
         &\leq (3+4\eps)\sum_{a=0}^{\gamma}\alpha_a \qquad \text{(by \Cref{eq:alpha-vs-beta})}\\
         &= (3+4\eps)d_T(x,r_{i+sk}) ,
    \end{split}
\end{equation}
as desired.
\end{proof}

We now bound the stretch of $\mathcal{T}$.

\begin{lemma}\label{lm:stretch-3ep} The stretch of $\mathcal{T}$ is $3+12\eps$ when $\eps\in (0,\frac18)$.
\end{lemma}
\begin{proof}
    We continue analyzing the stretch between two terminals $x$ and $y$. By \Cref{clm:xy-vs-xroot} and \Cref{lm:dist-x-r_isk}, we have
    \begin{equation*}
        \begin{split}
        d_T(x,y)&\geq \frac{(1-\eps)}{3+4\eps}\left(d_{T_i}(x,C(r_{i+sk})) + d_{T_i}(y,C(r_{i+sk}))\right) - 2\eps\\
        &\geq \frac{(1-\eps)}{3+4\eps}d_{T_i}(x,y) - 2\eps\\
        &\geq \frac{(1-\eps)}{3+4\eps}d_{T_i}(x,y) - \eps d_T(x,y) \quad \text{(as $d_T(x,y) \geq 8$).}
        \end{split}
    \end{equation*}
    This gives
    \begin{equation*}
        \begin{split}
            d_{T_i}(x,y)  \leq \frac{(1+\eps)(3+4\eps)}{1-\eps}d_T(x,y)\leq (3+12\eps)d_T(x,y)
        \end{split}
    \end{equation*}
    when $\eps \leq 1/8$.
\end{proof}

\subsection{Stretch $2+\eps$}\label{sec:dist-2e-cover}

The construction in the previous section has stretch at most $3+\eps$ due to the sum $2\sum_{a = 0}^t \beta_a$, called  \EMPH{$\beta$-sum}, in \Cref{eq:stretch-3}; each $\beta_a$ is approximately $\alpha_a$. Our idea is to reduce the contribution of the $\beta$-sum by doing an even more fine-grained bucketing of each $\bB_i$ modulo $p$ for $p \approx 1/\eps$ in \Cref{eq:k-p-def}. (So far, we have not really used $p$.) The key idea is that the fined-grained bucketing allows us to bound: $2\sum_{a = 0}^t \beta_a \approx \sum_{a=0}^t\alpha_a + \frac{1}{p}(\sum_{a = 0}^t \beta_a)$ which is approximately $(1+\eps)\sum_{a=0}^t\alpha_a$ for $p \approx 1/\eps$. Plugging this into \Cref{eq:stretch-3}, we get an improved version of \Cref{lm:dist-x-r_isk}, where the stretch is $2+O(\eps)$; this ultimately leads to stretch $2+O(\eps)$ in our non-Steiner tree cover.

\paragraph{Cover Construction.} 
Recall that $p = 1/\eps+1$. We construct $k$ buckets of chops $\bB_0, \ldots, \bB_{k-1}$ as described in \Cref{eq:bucket3-def}. We then further partition each bucket $\bB_i$ for every $i\in [0,k-1]$ into $p$ buckets $\bB_{i,0},\bB_{i,1}, \ldots, \bB_{i,p-1}$ as follows. For each $j\in [0,p-1]$

\begin{equation}
    \bB_{i,j} = \{\cL_{i+ (j+sp)\cdot k}: s\geq 0\} .
\end{equation}

We then obtain a set of $k\cdot p$ buckets of the form $\bB_{i,j}$ where $i\in [0,k-1], j = [0,p-1]$.   For each bucket $\bB_{i,j}$, we construct a tree $T_{i,j}$ in the same way we construct $T_i$ in the previous section. Our final tree cover is $\mathcal{T} = \{T_{i,j}: i\in [0,k-1], j\in [0,p-1]\}$

\begin{quote}
    \textbf{Constructing $T_{i,j}$.} Consider $\eps$-chops in $\bB_{i,j}$ from lower levels to higher levels. Let  $X$ be an  $\eps$-chop at level ${i+(j+sp)k}$  for some integers  $s\geq 0, j\geq 0$; assume for now that $s \geq 1$. Let $x$ be the root of $X$.  Let $Y$ the $\eps$-chop at level ${i+(j + (s-1)p)k}$ that is the child of $X$ in $\bB_{i,j}$.   We then add terminal $C(x)$ to $T_{i,j}$, and connect $C(x)$ to $C(y)$ with an edge. If $s = 0$, then we simply add a terminal $C(x)$ corresponding to the root of every tree $X$ in the chop at level $i+j$. Finally, we connect the roots corresponding to $s = 0$ to form a tree. We set of the weight of every edge $(t_1,t_2)$ in $T_{i.j}$ to be $d_T(t_1,t_2)$.
\end{quote}

\begin{figure}[h!]
    \centering
    \includegraphics[width=\textwidth]{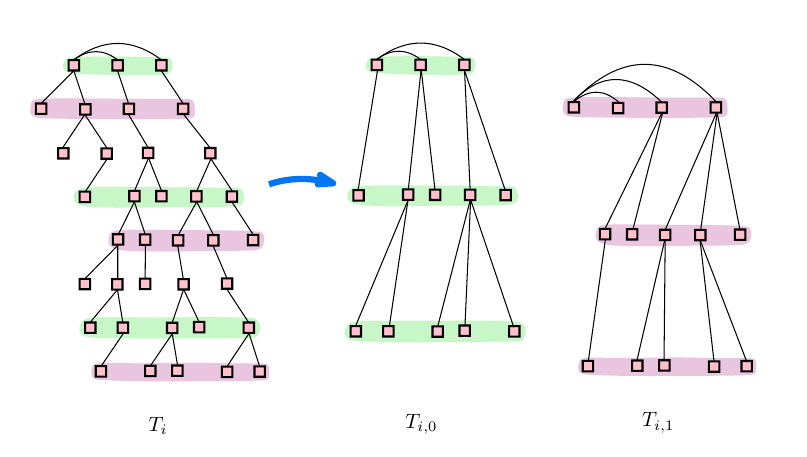}
    \caption{Construct two trees $T_{i,0}$ and $T_{i,1}$ from $T_{i}$ by skipping $3$ levels; here $p=3$.}
    \label{fig:skipConn}
\end{figure}

One can think of the construction of $T_{i,j}$ as skipping connections by exactly $p$ consecutive levels in the tree $T_i$ constructed in the previous section; see \Cref{fig:skipConn}. (Again, one terminal could have multiple copies in a tree of the cover, which could then be resolved by contraction.) By \Cref{lm:terminal-consecutive}, every terminal will appear in $T_{i,j}$. We now focus on the stretch analysis.

\paragraph{Stretch Analysis.} 
We consider any two terminals $x,y$ and $z = \lca(x,y)$. There must be some chop  $\cL_{i+(j+sp)k}$  such that $z$ belongs to the chop for some $s\geq 0$.  To simplify the presentation, w.l.o.g., we assume that $i = 0, j = 0, s = 0$. Thus, the chop containing $z$ is $\cL_{0}$.  Let $r_{0}$ be the root of the subtree containing $z$. \Cref{clm:xy-vs-xroot} remains true here, with $r_{0}$ in place of $r_{i+sk}$.

\begin{claim}\label{clm:xy-vs-xroot-2} $d_T(x,y)\geq (1-\eps)d_T(x,r_{0}) + (1-\eps)d_T(y,r_{0}) - 2\eps$.
\end{claim}

\begin{figure}[h!]
    \centering
    \includegraphics[width=\textwidth]{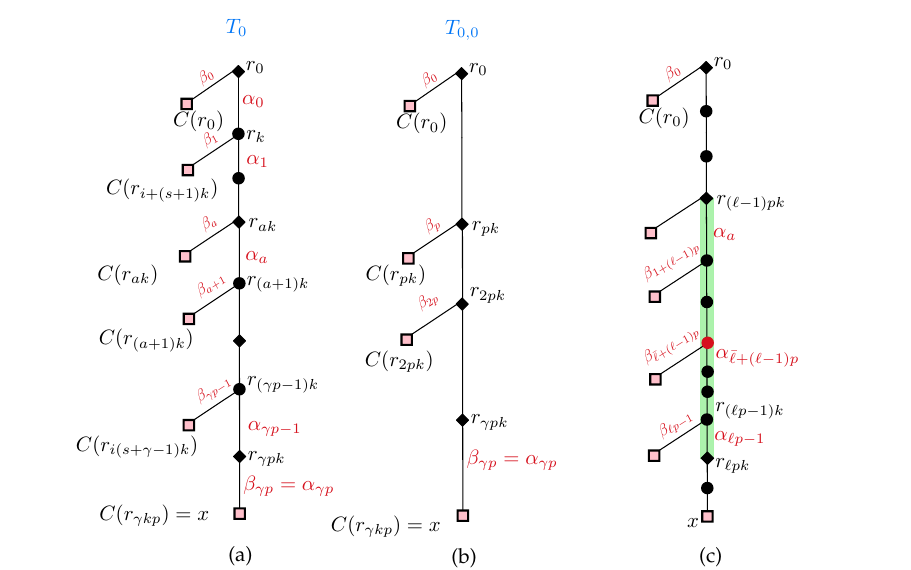}
    \caption{(a) In $T_0$, each root $r_{ak}$ will be replaced by the corresponding closest terminal; (b) the roots of the $\eps$-chops in the construction $T_{0,0}$, each root will also be replaced with a corresponding closest interval; (c) the root corresponding to $\Bar{\ell}$ is highlighted red.}
    \label{fig:dist2}
\end{figure}

We now show an analogous version of \Cref{lm:dist-x-r_isk}.

\begin{lemma}\label{lm:dist-x-r_isk-2} $d_{T_{0,0}}(C(r_{0}), x)\leq (2+2\eps)d_T(r_{0}, x)$. Similarly, $d_{T_{0,0}}(C(r_{0}), y)\leq (2+2\eps)d_T(r_{0}, y)$.
\end{lemma}
\begin{proof}
    We focus on $d_{T_{0,0}}(C(r_{0}), x)$.  Let  $r_{\gamma p k}$ for some integer $\gamma \geq 0$ be the root of a tree in a chop in $\bB_{0,0}$ closest to $x$ such that  $r_{\gamma p k}$ is on the path from $r_0$ to $x$; see \Cref{fig:dist2}(b). Let $r_{k}, r_{2k},\ldots, r_{\gamma p k-1}$ be all the roots of consecutive chops in $\bB_{0}$ (not $\bB_{0,0}$) that are on the path from $r_0$ to $r_{\gamma pk}$; see \Cref{fig:dist2}(a). We note that $C(r_{\gamma pk}) = x$ and $C(r_0), C(r_{pk}), C(r_{2pk}),\ldots,
    C(r_{\gamma pk})$ is a path from $C(r_0)$ to $x$ in $T_{0,0}$. If $\gamma = 0$, then $d_{T_{0,0}}(x,C(r_0)) = 0$ and the lemma trivially holds. We consider the case where $\gamma\geq 1$.

    For every $a \in [0,\gamma\cdot p]$, let $\beta_{a} = d_T(r_{a k}, C(r_{a k}))$ and $\alpha_{a} = d_T(r_{a k}, r_{(a+1)k})$, with $r_{(\gamma p+1)k}$ defined to be $x$ so that $d_T(r_{\gamma p k}, r_{(\gamma p+1)k})$ is well-defined. Note that $\alpha_{\gamma p} = \beta_{\gamma p}$. By \Cref{lm:dist-chops-modk}, we have:
    \begin{equation}\label{eq:alpha-beta-2}
        \begin{split}
            \beta_a &\leq (1+2\eps)\alpha_{a} \qquad \forall a \in [0,\gamma p]\\
            \beta_{a+1} &\leq \alpha_a + \beta_a \qquad \forall a \in [0,\gamma p-1].
        \end{split}
    \end{equation}

    Next, we observe that
    \begin{equation}\label{eq:dist-r0-x}
           d_{T_{0,0}}(C(r_{0}), x)  \leq \sum_{a = 0}^{\gamma p-1}\alpha_a + \beta_ 0 + 2\sum_{\ell=1}^{\gamma-1}\beta_{\ell p} \qquad \text{ and } \qquad  \sum_{a = 0}^{\gamma p}\alpha_a = d_T(r_0,x).
    \end{equation}
    
For every $\ell \in [1,\gamma]$, we define $\Bar{\ell} \in [1,p-1]$ such that $\beta_{\Bar{\ell} + (\ell-1)p} = \min\{\beta_{1+(\ell-1)p}, \beta_{2+(\ell-1)p}, \ldots, \beta_{\ell p-1}\}$. See \Cref{fig:dist2}(c).  We have two claims:

\begin{claim}\label{clm:beta-ellp} $\beta_{\ell p} \leq \beta_{\Bar{\ell} + (\ell-1)p} + \sum_{a = \Bar{\ell} + (\ell-1)p}^{\ell p-1} \alpha_{a}$
\end{claim}
    By applying \Cref{eq:alpha-beta-2} repeatedly, we have:
    \begin{equation*}
        \begin{split}
            \beta_{\ell p} &\leq \beta_{\ell p - 1} + \alpha_{\ell p-1}\\
        &\leq  \beta_{\ell p - 2}  +  \alpha_{\ell p-2} + \alpha_{\ell p-1}\\
        &\leq \beta_{\Bar{\ell} + (\ell-1)p} +  \sum_{a = \Bar{\ell} + (\ell-1)p}^{\ell p-1} \alpha_{a} ,
        \end{split}
    \end{equation*}
implying \Cref{clm:beta-ellp}.

\begin{claim}\label{clm:beta-ellbar} $\beta_{\Bar{\ell}+(\ell-1)p} \leq \frac{1+\eps}{p-1}\sum_{a = 1+(\ell-1)p}^{\ell p-1}\alpha_{a}$.
\end{claim}
  By the definition of $\Bar{\ell}$, we have:
    \begin{equation*}
        \beta_{\Bar{\ell} + (\ell-1)p} \leq \frac{1}{p-1}\sum_{a = 1+(\ell-1)p}^{\ell p-1}\beta_{a} \leq \frac{1+\eps}{p-1}\sum_{a = 1+(\ell-1)p}^{\ell p-1}\alpha_{a}
    \end{equation*}
    by \Cref{eq:alpha-beta-2}, implying \Cref{clm:beta-ellbar}.

We continue with bounding $d_{T_{0,0}}(x,C(r_0))$. By \Cref{clm:beta-ellp}, we have
\begin{equation}\label{eq:bet-sum-vs-alpha}
    \begin{split}
        \sum_{\ell=1}^{\gamma}\beta_{\ell p} &\leq  \sum_{\ell=1}^{\gamma}\left(\beta_{\Bar{\ell} + (\ell-1)p} + \sum_{a = \Bar{\ell} + (\ell-1)p}^{\ell p-1} \alpha_{a}\right)\\
        &\leq \sum_{\ell=1}^{\gamma}\left(\beta_{\Bar{\ell} + (\ell-1)p} + \sum_{a = 1+ (\ell-1)p}^{\ell p-1} \alpha_{a}\right) \qquad \text{(since $\Bar{\ell}\geq 1$)}\\
        &\leq  \sum_{\ell=1}^{\gamma}\sum_{a = 1+ (\ell-1)p}^{\ell p-1} (1+\frac{1+\eps}{p-1})\alpha_{a} \qquad \text{(by \Cref{clm:beta-ellbar})} \\
    \end{split}
\end{equation}

By \Cref{eq:dist-r0-x}, we have
\begin{equation*}
    \begin{split}
         d_{T_{0,0}}(C(r_{0}), x)  &\leq \sum_{a = 0}^{\gamma p-1}\alpha_a + \sum_{\ell=0}^{\gamma}\beta_{\ell p} + \sum_{\ell=1}^{\gamma}\beta_{\ell p} \\
         &\leq  \sum_{a = 0}^{\gamma p-1}\alpha_a  + \sum_{\ell=0}^{\gamma}(1+\eps)\alpha_{\ell p} + \sum_{\ell=1}^{\gamma}\beta_{\ell p} \qquad \text{(by \Cref{eq:alpha-beta-2})}\\
    &\leq  \sum_{a = 0}^{\gamma p-1}\alpha_a  + \sum_{\ell=0}^{\gamma}(1+\eps)\alpha_{\ell p} +   \sum_{\ell=1}^{\gamma}\sum_{a = 1+ (\ell-1)p}^{\ell p-1} (1+\frac{1+\eps}{p-1})\alpha_{a}  \qquad \text{(by \Cref{eq:bet-sum-vs-alpha})}\\   &\leq  \sum_{a = 0}^{\gamma p-1}\alpha_a  + (1+2\eps)\left(\sum_{\ell=0}^{\gamma}\alpha_{\ell p} +   \sum_{\ell=1}^{\gamma}\sum_{a = 1+ (\ell-1)p}^{\ell p-1}\alpha_{a}\right)  \qquad \text{(since $p=1/\eps+1$ and $\eps\leq 1$)}\\    
    &\leq (2+2\eps)\sum_{a = 0}^{\gamma p}\alpha_a  \\
    &\leq  (2+2\eps) d_T(r_0,x) \qquad \text{(by \Cref{eq:dist-r0-x})}
    \end{split}
\end{equation*}
as desired.
\end{proof}

Now we could bound the stretch of $\mathcal{T}$.

\begin{lemma}\label{lm:stretch-2ep} The stretch of $\mathcal{T}$ is at most $2+10\eps$ when $\eps\in (0,1/8)$.
\end{lemma}
\begin{proof}
  By \Cref{clm:xy-vs-xroot-2} and \Cref{lm:dist-x-r_isk-2}, we have
    \begin{equation*}
        \begin{split}
        d_T(x,y)&\geq \frac{(1-\eps)}{2+2\eps}\left(d_{T_{0,0}}(x,C(r_{0})) + d_{T_{0,0}}(y,C(r_{0}))\right) - 2\eps\\
        &\geq \frac{(1-\eps)}{2+2\eps}d_{T_{0,0}}(x,y) - 2\eps\\
        &\geq \frac{(1-\eps)}{2+2\eps}d_{T_{0,0}}(x,y) - \eps d_T(x,y) \quad \text{(as $d_T(x,y) \geq 8$)} .
        \end{split}
    \end{equation*}
    This gives
    \begin{equation*}
        \begin{split}
            d_{T_{0,0}}(x,y)  \leq \frac{(1+\eps)(2+3\eps)}{1-\eps}d_T(x,y)\leq (2+10\eps)d_T(x,y)
        \end{split}
    \end{equation*}
    when $\eps \leq 1/8$.
\end{proof}

\subsection{Stretch $2$}\label{sec:dist-2-cover}

In this subsection, we construct a tree cover of stretch $2$ that has $O(\log n)$ trees. We say that a vertex $v$ is a \EMPH{centroid} of a tree $T$ if every connected component of $T\setminus \{v\}$ has at most $n/2$ vertices. Initially, $\mathcal{T} = \emptyset$. For each vertex $u$, we denote by $C(u)$ the closest terminal to $u$. (If $u$ is a terminal, then $C(u) = u$.) Our construction is recursive. 

\begin{enumerate}
    \item \textbf{Step 1.} If $T$ contains a single terminal (the base case), then we simply return a singleton tree.   Otherwise,  we find a centroid $v$ of $T$. Then we make a star $X_v$ with center $C(v)$, and for every terminal $u\in K\setminus \{C(v)\}$, we add an edge $(C(v), u)$ to $X_v$ and set the weight $w_{X_v}(C(v), u) = d_T(u,C(v))$. Then we add $X_v$ to $\mathcal{T}$.
    \item  \textbf{Step 2.} Let $\Bar{T}_1, \Bar{T}_2, \ldots, \Bar{T}_\kappa$ be all the connected components of $T\setminus \{v\}$. We recursively construct a non-Steiner tree cover $\Bar{\mathcal{T}}_j$ for each component $\Bar{T}_j$, where $j\in [\kappa]$. Let $s = \max_{j\in [\kappa]}|\Bar{\mathcal{T}}_j|$. By making duplicate copies if necessary, we assume that every cover $\Bar{\mathcal{T}}_j$ contains exactly $s$ trees, denoted by $\{Y^j_1,Y^j_2, \ldots, Y^j_s\}$. Then, we create $s$ trees $\{Z_1,Z_2\ldots, Z_s\}$: for each $a \in [s]$, the $a$-th tree $Z_{a}$ is formed by taking all the $a$-th trees $Y^1_a, Y^2_a,\ldots, Y^\kappa_a$, one from each tree cover; we then connect $Y^j_a$ to $Y^1_a$ for every $j\in \{2,\ldots, \kappa\}$ by adding an edge $(t_j,t_1)$ from an (arbitrary) terminal $t_j \in Y^j_a$ to an arbitrary terminal $t_1 \in Y^1_a$. By adding all the edges $(t_j,t_1)$, we effectively connect every $Y^j_a$ to $Y^1_a$, and finally get a tree $Z_{a}$. The weight of the edge $(t_j,t_1)$ is $w_{Z_a}(t_j,t_1) = d_T(t_j,t_1)$. We then add all the trees $Z_1,\ldots, Z_s$ to $\mathcal{T}$. 
\end{enumerate}

It follows directly from the construction that every tree in $\mathcal{T}$ is non-Steiner. Furthermore, for every tree $X\in \mathcal{T}$, every edge $(x,y)\in X$ has a weight $w_X(x,y) = d_T(x,y)$. Thus, by the triangle inequality, $X$ is dominating. It remains to bound the number of trees, as well as the stretch of $\mathcal{T}$.

\paragraph{Bounding the Number of Trees in $\mathcal{T}$.} 
Let $s(n)$ be the number of trees in $\mathcal{T}$ when applying the above algorithm to a tree $T$ with $n$ vertices. Then we have 
$$s(n)\leq s(n/2) + 1,$$ 
where the $+1$ term is due to the tree $X_v$ in Step~1, and $s(n/2)$ is an upper bound for the size of all the covers  $\Bar{\mathcal{T}}_1, \Bar{\mathcal{T}}_2, \ldots, \Bar{\mathcal{T}}_\kappa$, as each connected component of $T\setminus \{v\}$ has at most $n/2$ vertices. Solving the above recurrence gives $s(n) \le \lceil\log_2 n\rceil = O(\log n)$.

\paragraph{Analyzing the Stretch.} 
Let $x$ and $y$ be any two terminals in $T$. If $x$ and $y$ are in different components of $T\setminus \{v\}$. Then we have
\begin{equation}
    \begin{split}
        d_{X_v}(x,y) &= w_{X_v}(x,C(v)) + w_{X_v}(y,C(v))\\
        &= d_T(x,C(v)) + d_T(y,C(v))\\
        &\leq d_T(x,v) + d_T(v,C(v)) +  d_T(y,v)  + d_T(v,C(v)) \quad \text{(by triangle inequality)}\\
    &= d_T(x,y) + 2d_T(v,C(v))\\
    &\le d_T(x,y) + d_T(v,x) + d_T(v,y) \quad \text{(by definition, $d_T(v,C(v))\leq \min\{d_T(v,x) , d_T(v,y)\}$)}\\
    &=  2d_T(x,y),
    \end{split}
\end{equation}
implying that the stretch is at most $2$ for the pair $x,y$. If $x$ and $y$ are in the same component of $T\setminus \{v\}$, then we get stretch $2$ by induction.

\section{Steiner Spanners for Terminals in Planar Graphs}
\label{sec:Steiner}

Recall that a metric space $(X,d_X)$ is \emph{planar} if there exists an edge-weighted planar graph $G=(V,E,w)$ such that $X\subseteq V$ and $d_X$ is the shorted-path metric of $G$ restricted to $X$. 

\begin{restatable}{theorem}{SteinerSpannerPlanar}\label{thm:SteinerPlanar} Let $\eps \in (0,1)$ be a parameter. Let $T$ be a set of $n$ points (\emph{terminals}) in a planar metric. We can construct a Steiner $(1+\eps)$-spanner for $T$ with $O((n/\eps)\cdot \max\{\alpha(n), \log\eps^{-1}\}\cdot \log \eps^{-1})$ edges, where $\alpha(n)$ is the inverse Ackermann function. 
\end{restatable}

In this section, we prove \Cref{thm:steiner}. We start in \Cref{ssec:Steiner1} with reviewing a classical spanner construction based on a net trees, in planar metrics~\cite{LW21}. The problem of constructing $(1+\eps)$-spanners is reduced to additive spanners~\cite{LW21,ChangCLMST23} on each level of a net tree in \Cref{ssec:Steiner1}, and further to additive spanners in each subgraph in a cover decomposition in \Cref{ssec:Steiner3}. Finally in \Cref{ssec:Steiner4}, we construct the required additive spanners for bounded-diameter planar graphs using shortest path separators~\cite{FiltserL22,Klein02,Thorup04}
and tree shortcutting~\cite{AS87,BodlaenderTS94,Chazelle87a}.

\subsection{Net Trees Based Spanners}
\label{ssec:Steiner1}

Let $G=(V,E,w)$ be an edge-weighted planar graph, and let $T\subset V$ be a set of $n$ \emph{terminals}. In this section, we work with the planar metric $(T,d_G)$, where $d_G$ is the shortest path distance between the terminals in $G$. The \EMPH{aspect ratio} of the metric is the ratio of the maximum to the minimum distance between distinct vertices: $\rho=\max_{x,y\in T} d_G(x,y)/ \min_{x,y\in T, x\neq y} d_G(x,y)$.   Without loss of generality, we may assume that minimum distance between distinct terminals is 1, i.e., $\min_{x,y\in T, x\neq y} d_G(x,y)=1$. In particular, the maximum distance between terminals is $\diam_G(V)= \rho$.

An \EMPH{$\delta$-net} in a metric space $(X,d)$ is a subset $N\subset X$ such that for every $x\in X$ there exists $y\in N$ such that $d(x,y)\leq \delta$ (i.e., the closed balls of radius $\delta$ centered in $N$ cover $X$) and $\min_{x,y\in N,x\neq y} d(x,y)\geq \delta$ (i.e., the open balls of radius $\delta/2$ centered in $N$ are pairwise disjoint).

For a given $\eps\in (0,\frac14)$, we construct a hierarchy of nets on the terminals $T=N_0\supseteq N_1\supseteq \ldots \supseteq N_{\lceil \log_2 \rho\rceil}$, where $N_i$ is a $2^i$-net.
This hierarchy induces a \EMPH{net-tree} $\mathcal{T}$, where level $i$ of the tree is the net $N_i$, and the parent of a vertex $v\in N_i$ is $v\in N_{i+1}$ (if $v \in N_{i-1}$) or another vertex $u\in N_{i+1}$ such that $d_G(u,v)\leq 2^{i+1}$. Every vertex $v\in T$ has a unique ancestor in the net $N_i$, that we denote by $v^{(i)}\in N_i$. 
Using geometric series, we obtain 
\begin{equation}\label{eq:ancestor}
d_G(v,v^{(i)})\leq \sum_{j=1}^i d_G(v^{(j-1)},v^{(j)})\leq \sum_{j=1}^i 2^j < 2^{i+1}.
\end{equation}

We construct a spanner $H_{\rm net}$ for the planar metric $(T,d_G)$ as follows: At level $i=0$, we have $N_0=T$ and we connect every terminal $u\in T$ to all other terminal $v\in T$ such that $d_G(u,v)\leq \frac{18}{\eps}$. For every level $i\in \{1,2,\ldots ,\lceil \log_2\rho\rceil \}$, let $\Delta_i=2^i/\eps$; and connect every terminal $u\in N_i$ to all other terminal $v\in N_i$ such that 
\[
    8\, \Delta_i\leq d_G(u,v)\leq 19\,\Delta_i.
\] 
Using a standard proof by induction, we show that $H_{\rm net}$ is a $(1+\eps)$-spanner on $T$. 

\begin{lemma}\label{lem:net}
   For $\eps\in (0,\frac14)$, the graph $H_{\rm net}$ is a $(1+\eps)$-spanner for the metric $(T,d_G)$. 
\end{lemma}
\begin{proof}
Consider the $\binom{n}{2}$ point pairs $\{x,y\} \subset T$ sorted in nondecreasing order by distance $d_G(x,y)$. Let $\{x_j,y_j\}$ denote the $j$-th pair. We prove, by induction on $j$, that $d_{H_{\rm net}}(x_j,y_j)\leq (1+\eps)d_G(x_j,y_j)$. 

In the base case, $\{x_1,y_1\}$ is a closest pair in $G$, and we have $d_G(x_1,y_1)=1$ by assumption. The edge $xy$ was added to $H_{\rm net}$ at level 0, and so $d_{H_{\rm net}}(x_1,y_1)=d_G(x_1,y_1)$. In fact, the same argument holds for all pairs $\{x_j,y_j\}\subset V$ with $d_G(x_j,y_j)\leq \frac{18}{\eps}$. 

For the induction step, consider a pair $\{x_j,y_j\}\subset V$ with $d_G(x_j,y_j)>\frac{18}{\eps}$, and assume that $d_H(x,y)\leq (1+\eps)d_G(x,y)$ for all pairs $\{x,y\}\subset V$ such that $d_G(x,y)<d_G(x_j,y_j)$. Then $x_j$ and $y_j$ are not adjacent at level 0.
Since $d_G(x_j,y_j)\leq \diam_G(V)\leq \rho$, there exists $i\in \{1,2,\ldots ,\lceil \log_2\Delta\rceil \}$ such that 
\[
   9\,\Delta_i<d_G(x_j,y_j)\leq 18\, \Delta_i.
\]
Considering the ancestors of $x_j$ and $y_j$ at level $i$,  Equation~\eqref{eq:ancestor} and the triangle inequality yield 
\begin{align*}
d_G(x_j^{(i)},y_j^{(i)}) 
   &\leq  d_G(x_j^{(i)},x_j) +  d_G(x_j,y_j)+  d_G(y_j,y_j^{(i)})
    \leq 18\,\Delta_i + 2\cdot 2^{i+1}
    \leq \left(4+4\eps \right) \Delta_i 
    \leq 19\,\Delta_i,\\
d_G(x_j^{(i)},y_j^{(i)}) 
  &\geq d_G(x_j,y_j) - d_G(x_j^{(i)},x_j) -  d_G(y_j,y_j^{(i)})
   \geq 9\,\Delta_i -2\cdot 2^{i+1}
   \geq \left(2-4\eps\right) \Delta_i 
   \geq 8\, \Delta_i,
\end{align*} 
for $\eps\in (0,\frac14)$. Consequently, the edge $x_j^{(i)}y_j^{(i)}$ has been added to $H_{\rm net}$ at level $i$.

By the induction hypothesis, $H_{\rm net}$ contains paths $\pi(x_j,x_j^{(i)})$ and $\pi(x_j,x_j^{(i)})$ of length at most $(1+\eps)2^{i+1}$. 
Concatenate the path $\pi(x_j,x_j^{(i)})$, the edge $x_j^{(i)} y_j^{(i)}$, and the path  $\pi(y_j,y_j^{(i)})$. We obtain a path in $H_{\rm net}$ between $x_j$ and $y_j$ of length
\begin{align*}
w(\pi(x_j,x_j^{(i)})) + d_G(x_j^{(i)},y_j^{(i)}) + w(\pi(y_j,y_j^{(i)}))
    &\leq d_G(x_j^{(i)},y_j^{(i)}) + 2(1+\eps)2^{i+1}\\
   &\leq d_G(x_j^{(i)},x_j) + d_G(x_j,y_j) + d_G(y_j,y_j^{(i)}) + 4(1+\eps)2^i\\
    &\leq d_G(x_j,y_j) + 2\cdot 2^{i+1} +4(1+\eps)2^i\\
    &= d_G(x_j,y_j) + (8+4\eps) 2^i\\
    &\leq d_G(x_j,y_j) + (8+4\eps) \eps \Delta_i\\
    &\leq d_G(x_j,y_j) + (8+4\eps)\eps\cdot \frac19\, d_G(x_j,y_j)\\
    &= \left(1+ \frac{8+4\eps}{9}\cdot \eps\right) \cdot d_G(x_j,y_j)\\
    &\leq (1+\eps)\cdot d_G(x_j,y_j),
\end{align*}
as required. This completes the induction step, hence the entire proof.
\end{proof}

\paragraph{Reduction to Additive Spanners in Net-Trees}

We reduce the problem to additive spanners. Specifically, we show that \Cref{lem:AdditiveSpanner} below implies \Cref{thm:steiner}. The proof of  \Cref{lem:AdditiveSpanner} is presented in  \Cref{ssec:Steiner3,ssec:Steiner4}

\begin{restatable}{lemma}{AdditiveSpanner}
\label{lem:AdditiveSpanner}
For every $i\in \mathbb{N}$, there exists a spanner $H_i$ on $N_i$ such that 
\begin{enumerate}
\item\label{lem:AS1} for all $x,y\in N_i$, 
    if $d_G(x,y)=\Theta(\Delta_i)$, 
    then $d_{H_i}(x,y)\leq d_G(x,y)+\eps \Delta_i$, and
\item\label{lem:AS2} $|E(H_i)|\leq 
    O\Big(|N_i|\, \eps^{-1}\cdot \log(\eps^{-1}\alpha(n))\Big)$,
\end{enumerate}
where $\alpha(.)$ denotes the inverse Ackermann function.
\end{restatable}

\paragraph{Spanner construction.}
Let $H_i$ be the additive spanners provided by \Cref{lem:AdditiveSpanner} for each level $N_i$ of a net tree $\mathcal{T}$; and let $H=\bigcup_{i\in \mathbb{N}} H_i$.

\begin{lemma}\label{lem:H-spanner}
For $\eps\in (0,\frac14)$, the graph $H$ is a $(1+2\eps)$-spanner for the metric $(T,d_G)$. 
\end{lemma}
\begin{proof}
  Let $x,y\in T$ be a pair of terminals in the edge-weighted graph $G=(V,E,w)$. By Lemma~\ref{lem:net}, the net-based spanner $H_{\rm net}$ contains an path $\pi_{xy}=(x=v_0,v_1,v_2,\ldots , v_k=y)$ of weight at most $(1+\eps)d+_G(x,y)$.
  
 Each edge $e$ of $H_{\rm net}$ was added at some level of the net-tree $\mathcal{T}$. Recall that at level~0, we added edges of length in the range $[1,18/\eps]$. We can also partition the edges at level~0 into $O(\log \eps^{-1})$ subsets such that in each subset the ratio between the edge lengths is bounded by a constant: For every edge $e$ of $H_{\rm net}$, there is an index $i\in \{-\lceil \log_2 (18/\eps)\rceil, \ldots, \lceil \log_2\rho\rceil\}$ such that $8\Delta_i\leq w(e)\leq 19\Delta_i$.  In particular, for every $j\in\{1,\ldots , k\}$, there exists $i(j)\in \{-\lceil \log_2 (18/\eps)\rceil, \ldots, \lceil \log_2\rho\rceil\}$ such that $8\Delta_{i(j)}\leq d_G(v_{j-1},v_j)\leq 19\Delta_{i(j)}$.
  
By Lemma~\ref{lem:AdditiveSpanner1}, for every $j\in \{1,\ldots , k\}$, the additive spanner $H_{i(j)}$ contains a path $\pi(v_{j-1},v_j)$ from $v_{j-1}$ to $v_j$ of weight 
$w(\pi(v_{j-1},v_j))\leq d_G(v_{j-1},v_j)+\eps\Delta_{i(j)} \leq (1+\eps/8)\cdot d_G(v_{j-1},v_j)$.  The concatenation of the paths $\pi(v_0,v_1),\ldots, \pi(v_{k-1},v_k)$ is a path in $H$ of weight at most
\begin{align*}
 d_H(x,y)
 &\leq \sum_{j=1}^k w(\pi(v_{j-1},v_j))\\
 &\leq \sum_{j=1}^k \left(1+\frac{\eps}{8}\right)\cdot d_G(v_{j-1},v_j)\\
 &= \left(1+\frac{\eps}{8}\right) w(\pi_{xy})\\
 &\leq \left(1+\frac{\eps}{8}\right) (1+\eps)d_G(x,y)\\
 &=\left(1+\frac98\eps+\frac{\eps^2}{8}\right)d_G(x,y)\\
 &< (1+2\eps)d_G(x,y),
\end{align*}
for any $\eps\in (0,\frac14)$, as claimed.
\end{proof}

\subsection{Reduction to Planar Metrics of Bounded Diameter}
\label{ssec:Steiner3}

Consider the planar metric $(N_i,d_G)$ on a single level of the net-tree. In this section, we reduce the problem of construction of $H_i$ (claimed in Lemma~\Cref{lem:AdditiveSpanner}) to subspaces of $N_i$ of bounded diameter. 

\begin{definition}
A \EMPH{$(\beta,s,\Delta)$-sparse cover} for a graph $G$ is a collection  $\mathcal{C}=\{C_1,\ldots ,C_t\}$ of subgraphs of $G$ (called \emph{clusters}) such that 
\begin{enumerate}
    \item $\diam(C_j)\leq \Delta$;
    \item for every $v\in V(G)$, there is a cluster $C_j\in \mathcal{C}$ such that $B_G(v,\Delta/\beta)\subseteq C_j$ (that is, $C_j$ contains all vertices at distance at most $\Delta/\beta$ from $v$); and
    \item every $v\in N_i$ is contained in at most $s$ clusters (that is, $|\{j: v\in C_j\in \mathcal{C}\}|\leq s$).  
\end{enumerate}
\end{definition}

Sparse covers were introduced by Awerbuch and Peleg~\cite{AwerbuchP90b}. For planar graphs, Busch et al.~\cite{BuschLT14} showed that one can construct $(\beta,s,\Delta)$-sparse cover for any $\Delta$ with constant $\tau$ and $s$. 

\begin{theorem}[\cite{BuschLT14}]\label{thm:BLT14}
    There exist absolute constants $\beta,s\in O(1)$ such that for every $\Delta>0$ and every planar graphs, a $(\beta,s,\Delta)$-sparse cover can be constructed in polynomial time.  
\end{theorem}

Recall that for each level $i$ of the net tree, the minimum distance between any two points of the net $N_i$ is at least $2^i=\eps\cdot \Delta_i$. Using \Cref{thm:BLT14} with parameter $\Delta=20\beta \Delta_i$,
we obtain a $(\beta,s,\Delta_i)$-sparse cover $\mathcal{C}_i=\{C_1,\ldots ,C_t\}$ of the planar graph $G$. We only care about the clusters $C_j\in \mathcal{C}$ where $|N_i\cap C_j|\geq 2$.  

\subsection{Recursive Shortest-Path Separators}
\label{ssec:Steiner4}

Consider a single cluster, $C_j\in \mathcal{C}_i$, which is a planar graph with diameter $O(\Delta_i)=O(2^i/\eps)$, and recall the distance between any two net points in $N_i$ is at least $\eps \Delta_i$. In this section, we construct a $(1+\eps)$-spanner for $N_i\cap C_j$ when $|N_i\cap C_j|\geq 2$. 

\paragraph{Shortest Path Separators.}
We recursively partition $C_j$ along shortest path separators until each subgraph contains at most one net point in $N_i\cap C_j$. A \emph{balanced separator} (for short, \emph{separator}) of a graph $G$ is a set of vertices $S\subset V(G)$ such that every connected component of $G-S$ has at most $\frac23\cdot |V(G)|$ vertices. According to a celebrated result by Lipton and Tarjan~\cite{LT79}, every $n$-vertex planar graph admits a balanced separator of size $O(\sqrt{n})$  A recursive partition of planar (or minor-free) graphs along balanced separators is well-known powerful technique; see~~\cite{AwerbuchP90a,ChangCLMST23,ChangCLMST24,Filtser20,JiaLNRS05,KleinMS13,Thorup04} for examples. Goodrich~\cite{Goodrich95} noticed that one can choose balanced separators in planar graph as the vertices of a \emph{fundamental cycle}, which his composed of two shortest paths from a common endpoint such that the other two endpoints of the paths are adjacent in $G$ (if $G$ is a triangulation) or at least incident to a common face (in general). Such a balanced separator is called a \emph{shortest path separator}. Since can use a shortest path tree to recursively partition a planar graph~\cite{Klein02,Thorup04}, and maintain the additional property that the each subgraph in the recursion is bounded by $O(1)$ shortest paths. We use the terminology presented in \cite{FiltserL22}: 

\begin{definition}\label{def:RSPD}
Given an edge-weighted graph $G=(V,E,w)$, a vertex $r\in V$, and a parameter $\eta>0$, an \EMPH{$\eta$-rooted shortest path decomposition}
(for short, \EMPH{$\eta$-RSPD}) with root $r$, denoted by $\Phi$, is a binary tree with the following properties:

%\begin{enumerate}
%\item[(P1)]\label{pp:P1} 
%$\Phi$ has height $O(\log n)$ and $O(n)$ nodes, where $n=|V|$.
%\item[(P2)]\label{pp:P2} 
Each node $\alpha\in \Phi$ is associated with a subgraph $G_\alpha$ of $G$, called a \emph{piece}, such that:
    \begin{enumerate}
    \item[(P1)]\label{pp:P1} The subtree of $\Phi$ rooted at $\alpha$ 
    has height $O(\log |V(G_\alpha)|$.
    \item[(P2)]\label{pp:P2} For every piece $G_\alpha$, there is a set of \emph{boundary vertices} $Q_\alpha\subset V(G_\alpha)$ such that every path between a vertex $u\in V(G_\alpha)$ and $v\in V(G)\setminus V(G_\alpha)$ in $G$ contains a vertex in $Q_\alpha$.
    The vertices in $V(G_\alpha)\setminus Q_\alpha$ are called \emph{internal vertices} of $G_\alpha$. 
    \item[(P3)]\label{pp:P3} For every piece $G_\alpha$, all boundary vertices in $Q_\alpha$ are contained in at most $\eta$ shortest paths in $G$ with a common endpoint $r$.
    \item[(P4)]\label{pp:P4} If $\alpha$ is the root of $\Phi$, then $G-\alpha=G$; if $\alpha$ is a leaf of $\Phi$, then $G_\alpha$ has at most $\eta$ internal vertices. Otherwise, $\alpha$ is an internal node of $\Phi$ with exactly two children $\beta_1$ and $\beta_2$. It holds that $G_\alpha= G_{\beta_1}\cup G_{\beta_2}$ and $V(G_{\beta_1})\cap V(G_{\beta_1})\subseteq Q_{\beta_1}\cap Q_{\beta_2}$. 
    \end{enumerate}
%\end{enumerate}
\end{definition}

Let $G=(V,E,w)$ be an edge-weighted planar graph. We may assume that $G$ is a triangulation (i.e., an edge-maximal planar graph) by adding edges of sufficiently large weight (that do not change the shortest path distance). Thorup~\cite[Section~2.5]{Thorup04} showed that for a triangulated wedge-weighed planar graph with $n$ vertices, one can compute a an $\eta$-RSPD with $\eta=O(1)$ in $O(n\log n)$ time (an RSPD is called \emph{frame separator decomposition} in Thorup's paper).

Note that if $\alpha$ and $\beta$ are siblings in $\Phi$, then $G_\alpha$ and $G_\beta$ may share boundary vertices. That is, 
$V(G)=\bigcup \{V(G_\alpha) : \alpha\in \Phi \mbox{ \rm is a leaf node}\}$ is not a partition. In order to avoid duplication, for each net point $v\in N_i\cap G$, we specify a lowest node $\varphi(v)\in \Phi$ such that $v\in V(G_{\varphi(v)})$. For every node $\alpha\in \Phi$, let $N_\alpha$ denote the set of net points $v\in N_i$ such that $\varphi(v)$ is a descendant of $\alpha$ (possibly $\varphi(v)=\alpha)$. With this notation, 
$N_i=\bigcup \{N_\alpha: \alpha\in \Phi \mbox{ \rm is a leaf node}\}$ is a partition; and we have $|N_i|=\sum_{\mbox{\rm leaf } \alpha\in \Phi} |N_\alpha|$.

Filtser and Le~\cite{Filtser20} used the $\eta$-RSDP for an (exact) emulator for tree metrics with treewidth $O(\log \log n)$ and hop-diameter $O(\log \log n)$; not for Steiner spanners. However, they proved the following lemma. 

\begin{lemma}[Lemma~4 in~\cite{FiltserL22}]\label{lem:FL22}
Let $\alpha$ and $\beta$ be two nodes in $\Phi$. Let $P_{uv}$ be any path between two vertices $u$ and $v$ in $G$ such that $u\in Q_\alpha$ and $v\in Q_\beta$. 
Let $(\alpha= \lambda_1,\lambda_2,\ldots , \lambda_k=\beta)$ be a set of nodes in the unique path $\Phi[\alpha,\beta]$ such that $\lambda_{i+1}\in \Phi[\lambda_i,\beta]$ for any $1\leq i\leq k-1$. 
Then, there exists a sequence of vertices 
$(u = x_1,x_2,\ldots ,x_k = v)$ such that 
$x_i \in P_{uv}\cap Q_{\lambda_i}$ and 
$x_{i+1}\in P_{uv}\cap P[x_i,v]$ for any $1\leq i\leq k-1$.
\end{lemma}

\begin{remark}
Assume that $\Phi[\alpha,\beta]=(\alpha= \lambda_1,\lambda_2,\ldots , \lambda_k=\beta)$ is the (unique) path in $\Phi$ between $\alpha$ and $\beta$. Then Lemma~\ref{lem:FL22} states that the shortest path $P_{uv}$ contains some boundary vertices from $Q_{\lambda_1},\Q_{\lambda_2},\ldots ,Q_{\lambda_k}$ is this order (a vertex can belong to several boundary sets). However, it does \emph{not} say that the shortest path is contained in the union of these pieces, $\bigcup_{i=1}^k G_{\lambda_i}$. It is possible that $P_{uv}$ passes though some additional pieces---but this will not affect our construction.
\end{remark}

\paragraph{Shortcut Edges in RSPD.}
Let $\Phi$ be an $\eta$-RSPD for $G$ with a constant $\eta$. It is a binary tree of height $O(\log n)$, and so its diameter is $O(\log n)$. We augment $\Phi$ with \emph{shortcut} edges in order to reduce its diameter.

Chung and Garay~\cite{ChungG84} initiated the studied the minimum number of shortcut edges for paths and trees. Chazelle~\cite{Chazelle87a} showed that a $n$-vertex tree can be augmented with $m$ new edges to reduce the diameter to $O(\alpha(n,m))$, where $\alpha(n,m)$ is the two-parameter inverse Ackermann function. This bound is the best possible~\cite{Yao82}, and was later rediscovered several times~\cite{AS87,BodlaenderTS94}; see also~\cite{Bilo22a,FiltserL22,LMS22,Thorup97} for algorithmic aspects and generalizations. Alternatively, the diameter of a tree with $n$ vertices can be reduced to $2k$ by adding $O(n\alpha_k(n))$ edges, 
where $\alpha_k(n)$ is the inverse of a certain Ackermann-style function at the $\lfloor k/2\rfloor$th level of the primitive recursive hierarchy: Specifically, $\alpha_0(n) = \lceil n/2\rceil$, $\alpha_1(n) = \lceil \sqrt{n}\rceil$, $\alpha_2(n)=\lceil \log n\rceil$, $\alpha_3(n)=\lceil \log\log n\rceil$, $\alpha_4(n)=\log^* n$, etc..  Moreover, $\alpha_{2\alpha(n)+2}(n)\leq 4$, where $\alpha(n)$ is the one-parameter inverse Ackermann function, which is an extremely slowly growing function~\cite{AS87,LMS22,Solomon13}. 

We apply the above results to the $\eta$-RSPD $\Phi$ carefully. For our purposes, we distinguish between two types of shortcut edges: \EMPH{type~1} shortcut edges between internal nodes of $\Phi$; and \EMPH{type~2} shortcut edges between a leaf and an internal node. For our purposes (discussed below), a shortcut edge of type~1 costs roughly $O(\eps^{-2})$; and one of type~2 costs roughly $O(\eps^{-1})$. For this reason, we prefer shortcut edges of type~2. We add shortcut edges to $\Phi$ as follows.

\begin{figure}[h!]
    \centering
    \includegraphics[width=0.95\textwidth]{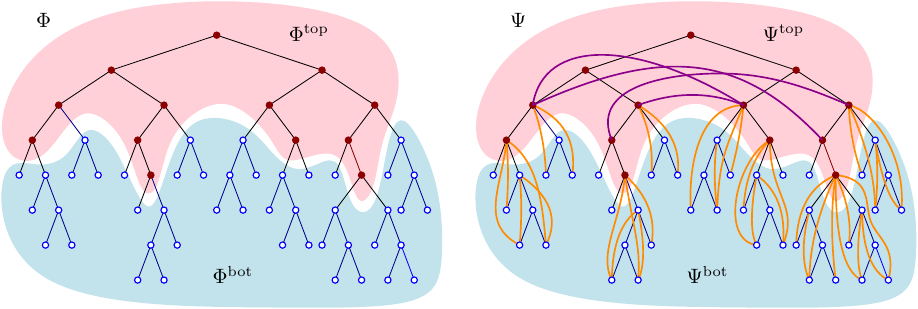}
    \caption{Left: a binary tree $\Phi$, with $\Phi^{\rm top}$ and $\Phi^{\rm bot}$. Right: The augmented graph $\Psi$, with $\Psi^{\rm top}$ and $\Psi^{\rm bot}$.}
    \label{fig:shortcuts}
\end{figure}

We decompose the tree $\Phi$ as follows. Let $\lambda$  be a parameter to be optimized later (we shall choose $\lambda=\eps^{-1} \alpha(n)/\log (\eps^{-1}\alpha(n))$). 
Let $\Phi^{\rm top}$ denote the subtree of $\Phi$ induced by all nodes
$\alpha$ with $|V(G_\alpha)|\geq \lambda$, and $\Phi^{\rm bot}$ the forest induced by all other nodes of $\Phi$ and the leaves of $\Phi^{\rm top}$. 
Then $\Phi^{\rm top}$ has $O(n/\lambda)$ nodes; and the height of (each tree in) $\Phi^{\rm bot}$ is $O(\log \lambda)$.
Now we augment $\Phi$ to a graph $\Psi$ with shortcut edges as follows (see Fig.~\ref{fig:shortcuts}):
\begin{enumerate}
    \item Reduce the diameter of $\Phi^{top}$ to $O(1)$~\cite{AS87,BodlaenderTS94,Chazelle87a};
    \item augment $\Phi^{\rm bot}$ by connecting every leaf of the forest $\Phi^{\rm bot}$ to all of its ancestors in $\Phi^{\rm bot}$. 
\end{enumerate}
Denote by $\Psi^{\rm top}$ and $\Psi^{\rm bot}$, resp., the subgraph of $\Psi$ induced by the vertices of $\Phi^{\rm top}$ and $\Phi^{\rm bot}$.

\begin{lemma}\label{lem:Psi}
The graph $\Psi^{\rm top}$ has $O(n\alpha(n)/\lambda)$ edges; and the forest $\Psi^{\rm bot}$ has $O(n\log \lambda)$ edges incident to its leaves.
\end{lemma}
\begin{proof} 
   Since $\Psi^{\rm top}$ is a binary tree on $O(n/\lambda)$ nodes, it has $O(n/\lambda)$ edges. The first augmentation phase adds $O(( n/\lambda)\alpha(n/\lambda)) \leq O(n \alpha(n)/\lambda)$ shortcut edges, and so $\Psi^{\rm top}$ has $O(n \alpha(n)/\lambda)$ edges. 

   The forest $\Phi^{\rm bot}$ has $O(n)$ leaves, each of which is incident to only one edge in $\Phi$. Since the height of $\Phi^{\rm bot}$ is $O(\log \lambda)$, the second augmentation step adds $O(n\log \lambda)$ new edges to every leaf. Overall, $\Psi^{\rm bot}$ has $O(n\log \lambda)$ edges incident to leaves.
\end{proof}

\paragraph{Portals along Shortest Path Separators.}
Recall that every node $\alpha$ corresponds to a piece $G_\alpha$, 
and the boundary vertices $Q_\alpha$ of $G_\alpha$ all lie in $\eta=\O(1)$ shortest paths of $G$. 

\begin{figure}[h!]
    \centering
    \includegraphics[width=0.8\textwidth]{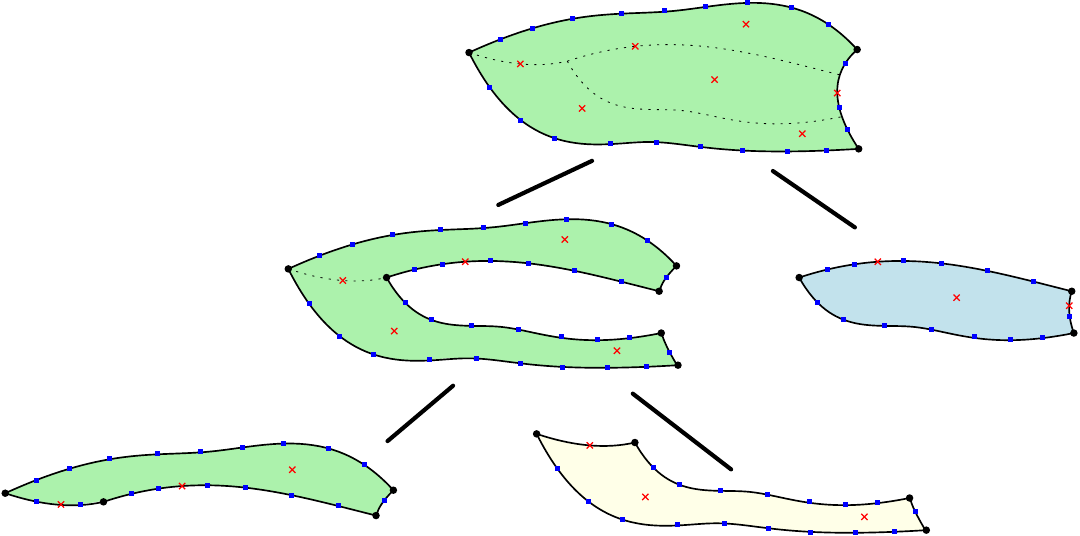}
    \caption{A recursive partition of a plane graph along shortest paths into pieces, until there are at most $\eta=O(1)$ terminals in the interior of each piece. Terminals (red crosses) and portals (blue squares).}
    \label{fig:portals}
\end{figure}

We place Steiner points, that we call \emph{portals}, along the shortest path in $Q_\alpha$ as follows; see Fig.~\ref{fig:portals}. Let $\alpha$ be a node of $\Phi$, and suppose that the vertices in $Q_\alpha$ lie in the shortest paths $P_1,P_2,\ldots ,P_\eta$ in $G$. The length of each path is at most $\diam(G)\leq \Delta_i$. For each $j$, $1\leq j\leq \eta$, we place portals at the two endpoint of $P_j$, and recursively place portals at internal nodes until any two consecutive portals at at distance at most $\frac{\eps}{10}\, \Delta_1$ apart or are adjacent along $P_j$.  Let $S_\alpha$ denote the set of portals (i.e., Steiner points) over all $\eta$ paths. It follows that we place $O(\eps^{-1})$ portals along each shortest path, and so $|S_\alpha|\leq O(\eta\cdot \eps^{-1}) = O(\eps^{-1})$.  

We can now define the Steiner spanner $H_{i,j}$ for $N_i\cap C_j$. Let the vertex set of $H_i$ be $N_i\cup \bigcup_{\alpha\in \Phi} S_\alpha$, that is, the net points in $N_i\cap C_j$ and all portals defined above. We add the following edges to $H_{i,j}$, each with the same weight as in $G$: 
\begin{enumerate}
    \item For every edge $\alpha\beta$ in $\Psi^{\rm top}$, add a complete bipartite graph between the portals $S_\alpha$ and $S_\beta$;
    \item for every edge $\alpha\beta$ of $\Psi$, where $\alpha$ is a leaf of $\Psi^{\rm bot}$, add a complete bipartite graph between $N_\alpha$ and $S_\beta$;
    \item for every leaf $\alpha$, add a complete graph among 
    its internal net vertices $N_i\cap (G_\alpha\setminus Q_\alpha)$ and a complete bipartite graph between $N_\alpha$ and $S_\alpha$.
\end{enumerate}

We are now ready to prove \Cref{lem:AdditiveSpanner}. 
\begin{lemma}\label{lem:AdditiveSpanner1}
    We have $|E(H_{i,j})|\leq O\Big(|N_i\cap C_j|\cdot  \eps^{-1}
    \log(\eps^{-1}\alpha(n))\Big)$,
\end{lemma}
\begin{proof}
By \Cref{lem:Psi}, the graph $\Psi^{\rm top}$ has $O(|N_i\cap C_j|\, \alpha(n)/\lambda)$ edges, and for each such edge we add a complete bipartite graph with $O(\eps^{-2})$ edges to $H_i$. This contributes  $O(\eps^{-2}\, |N_i\cap C_j|\, \alpha(n)/\lambda)$ edges to $H_i$.
By \Cref{lem:Psi}, the graph $\Psi$ has $O(|N_i\cap C_j|\log \lambda)$ edges incident to leaves of $\Psi^{\rm bot}$, and for each such edge we add a star with $O(\eps^{-1})$ edges to $H_{i,j}$. This contributes  $O(\eps^{-1} |N_i\cap C_j| \log\lambda)$ such edges to $H_{i,j}$.   

We choose $\lambda:=\eps^{-1} \alpha(n)/\log (\eps^{-1}\alpha(n))$ to balance the above two contributions. They both amount to 
$O\Big(\eps^{-1} |N_i\cap C_j| \log (\eps^{-1} \alpha(n))\Big)$.

Finally, each leaf node $\alpha\in \Phi$ has at most $\eta=O(1)$ internal vertices. The complete graphs among internal net vertices contribute at most $\eta\cdot |N_i\cap C_j|=O(|N_i\cap C_j|)$ edges
contribute a total of $O(\eta^2\cdot |N_i\cap C_j) = O(|N_i\cap C_j)$ edges.
\end{proof}

\begin{lemma}\label{lem:AdditiveSpanner2}
For all $x,y\in N_i\cap C_j$, if $d_G(x,y)=\Theta(\Delta_i)$, then $d_{H_{i,j}}(x,y)\leq d_G(x,y)+\eps \Delta_i$.
\end{lemma}
\begin{proof}
Let $x,y\in N_i$ such that $d_G(x,y)=\Theta(\Delta_i)$.
Then $x\neq y$, and the $\eta$-RSPC has two leaves $\varphi(x),\varphi(y)\in \Phi$ such that $x\in N_{\varphi(x)}$ and $y\in N_{\varphi(y)}$. We distinguish between three cases:

\paragraph{Case~1: $\varphi(x)=\varphi(y)$.} If both $x$ and $y$ are internal vertices of $G_{\varphi(\alpha)}$, then the complete graph on $N_\alpha\cap (G_{\varphi(\alpha)}\setminus Q_{\varphi(\alpha)})$  contains the edge $xy$. Otherwise we may assume w.l.o.g.\ that $x$ is a boundary vertex, that is, $x\in Q_{\varphi(\alpha)}$. Then 
$x$ lies on a shortest path in $Q_{\varphi(\alpha)}$, and there is a portal $s\in S_{\varphi(\alpha)}$ such that $d_G(x,s)\leq \frac{\eps}{10}\Delta_i$. The complete bipartite graph between $N_{\varphi(\alpha)}$ and $S_{\varphi(\alpha)}$ contains the edge $sy$.
Consequently,
\begin{align*}
    d_{H_{i,j}}(x,y)
    &\leq d_{H_i}(x,s)+ d_{H_i}(s,y) 
       = d_{G}(x,s)+ d_{G}(s,y)\\
     &\leq \Big(d_{G}(x,y) + d_{G}(s,y)\Big) + d_G(s,y) 
       =d_{G}(x,y) + 2d_{G}(s,y)\\
      & \leq d_{G}(x,y) + 2\cdot \frac{\eps}{10}\Delta_i
    <  d_{G}(x,y) + \eps \Delta_i,
\end{align*}
as required.

\paragraph{Case~2: $\varphi(x)\neq\varphi(y)$ and $\lca\{\varphi(x),\varphi(y)\}$ is in $\Phi^{\rm bot}$.}
By construction $\Psi$ contains the path $\Psi[\varphi(x),\lca\{\varphi(x),\varphi(y)\}, \varphi(y)]$. By Lemma~\ref{lem:FL22}, the shortest path $P_{xy}$ in $G$ contains a sequence of vertices $(x=x_0,x_1,x_2=y)$ in this order such that $x_1\in Q_{\lca\{\varphi(x),\varphi(y)\}}$. The construction of portals ensures that there exists a portal $s_1\in S_{\lca\{\varphi(x),\varphi(y)\}}$ with $d_G(x_1,s_)\leq \eps \Delta_i$. Consequently, 
\begin{align*}
    d_{H_{i,j}}(x_y) 
    &\leq d_{H_i}(x,s_1) + d_{H_i}(s_1,y) \\
    & = d_{G}(x,s_1) + d_{G}(s_1,y) \\
    &\leq \Big(d_{G}(x,s_1) + d_G(s_1,x_1)\Big)+ 
        \Big(d_G(x_1,s_1)+d_{G}(s_1,y)\Big) \\
    & \leq d_G(x,x_1) + d_G(x_1,y) + 2\cdot \frac{\eps}{10}\cdot \Delta_i \\
    &< d_G(x,y) + \eps\cdot \Delta_i,
\end{align*}
as required.

\paragraph{Case~3: $\varphi(x)\neq\varphi(y)$ and $\lca\{\varphi(x),\varphi(y)\}$ is not in $\Phi^{\rm bot}$.}
Nodes $\varphi(x)$ and $\varphi(y)$ have ancestors $\alpha$ and $\beta$, resp., that are leaves in $\Phi^{\rm top}$. Furthermore, $\alpha\neq \beta$ or else we would be in Case~2. Due to the shortcut edges, the distance between $\alpha$ and $\beta$ is at most 4 in $\Psi^{\rm top}$. 
Consequently, $\Psi$ contains a path $\Psi[\varphi(x),\varphi(y)]=(\varphi(x)=\lambda_0,\lambda_1,\ldots , \lambda_k=\varphi(y))$ of length at most $k\leq 6$. By Lemma~\ref{lem:FL22}, the shortest path $P_{xy}$ in $G$ contains a sequence of vertices $(x=x_0,x_1,\ldots , x_k=y)$ in this order such that $x_j\in Q_{\lambda_j}$.
The construction of portals ensures that there exist portals $s_j\in S_{\lambda_j}$ with $d_G(x_j,s_j)\leq \eps \Delta_i$ for all $j\in\{1,\ldots ,k-1\}$.
The construction of $H_{i,j}$ guarantees that the edges $xs_1$, $s_1s_2,\ldots ,s_{k-2}s_{k-1},s_{k-1}y$ are present in $H_i$. 
Consequently, 
\begin{align*}
    d_{H_{i,j}}(x_y) 
    &\leq d_{H_i}(x,s_1) + d_{H_i}(s_1,s_2) + \ldots  + d_{H_i}(s_{k-2},s_{k-1}) + d_{H_i}(s_{k-1},y) \\
    & = d_{G}(x,s_1) + d_{G}(s_1,s_2) + \ldots  + d_{G}(s_{k-2},s_{k-1}) + d_{G}(s_{k-1},y) \\
    &\leq \Big(d_{G}(x,s_1) + d_G(s_1,x_1)\Big)+ 
        \Big(d_G(x_1,s_1)+d_{G}(s_1,s_2) +d_G(s_2,x_2)\Big) + \ldots  + \Big(d_G(x_{k-1},s_{k-1}) + d_{G}(s_{k-1},y)\Big) \\
    & \leq \sum_{j=1}^k d_G(x_{i-1},x_i) + 2(k-1)\cdot \frac{\eps}{10}\cdot \Delta_i \\
    &\leq d_G(x,y) + \eps\cdot \Delta_i,
\end{align*}
as required.
\end{proof}

The combination of \Cref{lem:AdditiveSpanner1,lem:AdditiveSpanner2} implies \Cref{lem:AdditiveSpanner}, that we restate for convenience: 

\AdditiveSpanner*
\begin{proof}
At every level $i\in \mathbb{N}$, we have constructed a 
 $(\beta,s,\Delta_i)$-sparse cover $\mathcal{C}_i=(C_1,\ldots , C_{t(i)})$ of $G$; and for every $j\in \{1,\ldots, t\}$, we have constructed and an additive spanner $H_{i,j}$ for $N_i\cap C_j$. 
Let $H_i=\bigcup_{j=1}^t H_{i,j}$. We claim that
\begin{enumerate}
\item \textbf{(stretch condition)} for all $x,y\in N_i$, 
    if $\Delta_i \leq d_G(x,y)\leq \frac{\Delta_i}{\beta}$, 
    then $d_{H_i}(x,y)\leq d_G(x,y)+\eps \Delta_i$, and
\item  \textbf{(size condition)} $|E(H_i)|\leq 
    O\Big(|N_i|\cdot \eps^{-1} \log(\eps^{-1}\alpha(n))\Big)$,
\end{enumerate}
where $\alpha(.)$ denotes the inverse Ackermann function.

\textbf{Stretch analysis.} Let $x,y\in N_i$ such that $\Delta_i \leq d_G(x,y)\leq \frac{\Delta_i}{\beta}$. By the definition of $(\beta,s,\Delta_i)$-sparse covers, there exists a cluster $C_j\in \mathcal{C}_i$ such that $B_G(x,\Delta/\beta)\subseteq C_j$. Then $x,y\in C_j$. 
By \Cref{lem:AdditiveSpanner2}, we have $d_{H_i}(x,y)\leq d_{H_i,j}(x,y)\leq d_G(x,y)+\eps \Delta_i$, as required.

\textbf{Size analysis.} By the definition of $(\beta,s,\Delta_i)$-sparse covers, each net point $v\in N_i$ is contained in at most $s=O(1)$ clusters in $\mathcal{C}_i$. Consequently, 
$\sum_{j=1}^t |N_i\cap C_j| \leq s\cdot |N_i| = O(|N_i|)$.
Summation of the bound in \Cref{lem:AdditiveSpanner1} now yields 
\[
 |E(H_i)|\leq \sum_{j=1}^t |E(H_{i,j})| 
         \leq  \sum_{j=1}^t  O\Big( |N_i\cap C_j|\cdot \eps^{-1}
    \log(\eps^{-1}\alpha(n))\Big) 
         \leq O\Big(|N_i|\cdot \eps^{-1}
    \log(\eps^{-1}\alpha(n))\Big),
\]
as required. 
\end{proof}

\subsection{Proof of \Cref{thm:SteinerPlanar}}
\label{ssec:SteinerPlanar}

We can now put the pieces of the puzzle together and prove Theorem~\ref{thm:SteinerPlanar}. Recall the definition of the graph $H$. For each level $N_i$ of a net tree $\mathcal{T}$, \Cref{lem:AdditiveSpanner} yields an additive spanner $H_i$; and we put $H=\bigcup_{i\in \mathbb{N}} H_i$.
We already know (Lemma~\ref{lem:H-spanner}) that $H$ is a $(1+2\eps)$-spanner for the metric $(T,d_G)$. It remains to bound the number of edges in $H$. 

\begin{lemma}\label{lem:H-size}
The graph $H$ has $O\Big(n\cdot \log(\eps^{-1}\alpha(n))\cdot \eps^{-1} \log \eps^{-1}\Big)$ edges
\end{lemma}
\begin{proof}
We are given a set $T$ of $n$ terminals in an edge-weighted planar graph $G=(V,E,w)$. We defined $\lceil \log_2(18/\eps)\rceil+\lceil \log_2\varrho\rceil+1$ levels, 
constructed an $(\beta,s,\Delta_i)$-sparse cover $\mathcal{C}_i=(C_1,\ldots , C_{t(i)})$ on each level, and the created an additive spanner $H_{i,j}$ on the net points $N_i\cap C_j$ if $|N_i\cap C_j|\geq 2$ (\Cref{lem:AdditiveSpanner}). 
We may assume that $H_{i,j}$ is the empty graph when $|N_i\cap C_j|\leq 1$. Then by \Cref{lem:AdditiveSpanner1}, $H_{i,j}$ has 
\[
     \max\{0,|N_i\cap C_j|-1|\} 
    \cdot O\left (\eps^{-1}  \log(\eps^{-1}\alpha(n)) \right)
\]
edges for all $j\in \{1,2,\ldots,  t(i)\}$. To complete the proof of \Cref{lem:H-size}, it is  enough to show that 
\begin{equation}\label{eq:vanish}
    \sum_{i=-\lceil \log_2 (18/\eps)\rceil}^{\lceil \log_2\rho\rceil}
    \sum_{j=1}^{t(i)} \max\{0,|N_i\cap C_j|-1|\} 
    =O\left(\frac{n}{\eps}\right).
\end{equation}

The proof of Equation~\eqref{eq:vanish} is based on the following key observation: We have $\diam(N_i\cap C_j)\leq 20\beta \Delta_i = 20\beta 2^i/\eps$, and $N_i$ is a $2^i$-net (i.e., the minimum distance between net points in $N_i$ is $2^i$). This implies that at most one point of $N_i\cap C_j$ is present in $N_{i+M}$, where $M:=\lceil \log_2 20\beta/\eps \rceil$. That is,
\begin{equation}\label{eq:vvv}
  \max\{0,|N_i\cap C_j|-1|\}  
  \leq \Big |(N_i\setminus N_{i+M}) \cap C_j\Big| +1
  \leq 2\cdot \Big |(N_i\setminus N_{i+M}) \cap C_j\Big|.
\end{equation}

Based on this observation, we partition the levels into $M$ \EMPH{groups}: Specifically, for every $g\in \{0,1,\ldots ,M-1\}$, group~$g$ comprises all levels $i$ such that $i\equiv g\pmod M$. Recall that at each level $i$, every net point in $N_i$ is contained in at most $s=O(1)$ clusters of the $(\beta,s,\Delta_i)$-sparse cover $\mathcal{C}_i$.
Combined with Equation~\eqref{eq:vanish}, this gives
\[
    \sum_{j=1}^{t(j)} \max\{0,|N_i\cap C_j|-1|\}  
    \leq 2s\cdot \Big |N_i\setminus N_{i+M}\Big|.
\]
Summation over all levels in any group $g\in \{0,1,\ldots , M-1\}$ yields
\[
    \sum_{i\equiv g\pmod M}\sum_{j=1}^{t(i) } 
    \max\{0,|N_i\cap C_j|-1|\}  
    \leq 2s\cdot \sum_{i\equiv g\bmod M}
    \Big |N_i\setminus N_{i+M}\Big|
    \leq 2s\cdot |N_0| =O(n).
\]
Finally, summation over all $M=O(\eps^{-1})$ groups gives
\begin{align*}
    \sum_{i=-\lceil \log_2 (18/\eps)\rceil}^{\lceil \log_2\rho\rceil}
    \sum_{j=1}^{t(i)} \max\{0,|N_i\cap C_j|-1|\} 
    &= \sum_{g=0}^{M-1} \left(\sum_{i\equiv g\pmod M}\sum_{j=1}^{t(i) } 
    \max\{0,|N_i\cap C_j|-1|\}\right) \\
    &\leq \sum_{g=0}^{M-1} O(n)\
     = O\left(\frac{n}{\eps}\right),
\end{align*}
as required. 
\end{proof}
The combination of Lemmas~\ref{lem:H-spanner} and \ref{lem:H-size} readily implies Theorem~\ref{thm:steiner}.

\subsection{Generalization to Graphs of Bounded Genus}
\label{ssec:SteinerGeneral}

In this section, we generalize \Cref{thm:SteinerPlanar} to polyhedral metrics

\begin{restatable}{theorem}{Surface}\label{thm:SteinerSurface} 
Let $\eps \in (0,1)$ be a parameter. Let $T$ be a set of $n$ points (\emph{terminals}) in a polyhedral metric. We can construct a Steiner $(1+\eps)$-spanner for $T$ with $O((n/\eps)\cdot \log(\eps^{-1}\alpha(n))\cdot \log \eps^{-1})$ edges, where $\alpha(n)$ is the inverse Ackermann function.
\end{restatable}

Every step of the proof of \Cref{thm:SteinerPlanar} generalizes to graphs of bounded genus. We briefly sketch the key differences. 
In  \Cref{ssec:Steiner1}, the construction of net-tree based spanners and \Cref{lem:net,lem:AdditiveSpanner,lem:H-spanner} hold for any metric space $(X,d_X)$. 
In \Cref{ssec:Steiner3}, we used $(\beta,s,\Delta)$-sparse covers to reduce the problem to planar graphs of bounded diameter. Klein et al.~\cite{KleinPR93} constructed $(\beta,s,\Delta)$-sparse covers for minor-free classes of graphs, where $\beta$ and $s$ depend on $\mathcal{H}$
(i.e., $\beta$ and $s$ are constants for fixed $\mathcal{H}$). 
Busch et al.~\cite{BuschLT14} later constructed $(\beta,s,\Delta)$-sparse covers for planar graphs with better constants; and Abraham et al.~\cite{AbrahamGMW10} for $K_{r,r}$-free graphs for any $r\in \mathbb{N}$. 

In \Cref{ssec:Steiner4}, we used a $\eta$-RSPD ($\eta$-rooted shortest path decomposition) for planar graphs of bounded diameter, based on Thorup~\cite[Section~2.5]{Thorup04}.
For graphs of constant genus $g$, we first reduce this step to planar graphs at the expense of increasing the diameter by a factor of $O(g)$.~\footnote{We note that for graphs of bounded genus, Abraham and Gavoille~\cite{AbrahamG06} constructed a weaker decomposition than $\eta$-RSPD: Specifically, property~(P3) of $\eta$-RSPD stipulates that for every node $\alpha\in \Phi$, the boundary vertices $Q_\alpha$ are contained in at most $\eta$ shortest paths of $G_\alpha$. In~\cite{AbrahamG06}, the boundary vertices are covered in at most $\eta$ \emph{successive} shortest paths, that is, $Q_\alpha$ is contained in paths $\gamma_1,\ldots , \gamma_k$ where $k\leq \eta$, and $\gamma_i$ is a shortest path in the graph $G_\alpha\setminus \bigcup_{j<i}\gamma_j$. Unfortunately, the removal of one or more paths may increase the diameter---and is unsuitable for our purposes.}
A \EMPH{cut graph} is a graph $C$ embedded on a surface $S$ such that $S\setminus C$ is homeomorphic to a closed disk~\cite{ECdV17}. 
It is NP-hard to find the \emph{shortest} cut subgraph in a given graph embedded in an oriented surface~\cite{EricksonH04,CohenAddadVMM21}. However, a cut graph has a very simple structure: It consists of a tree $T$ with $g$ cross edges, and can be computed in $O(n)$ time~\cite{Eppstein03,EricksonH04,GilbertHT84}. If we start with a shortest path tree $T$, we obtain a cut graph $C$ that consists of $O(g)$ pairwise noncrossing shortest paths. When we cut the surface $S$ (and the graph $G$) along $C$, the edges and vertices on $C$ are duplicated. If we start with a graph $G$ of genus $g$, we obtain a \emph{planar} graph $G'$.
\begin{lemma} \label{lem:cut}
\begin{enumerate}
    \item[]
    \item\label{lem:cut1} Every shortest path in $G'$ is the union of $O(g)$ shortest paths in $G$, and
    \item\label{lem:cut2} every shortest path in $G$ is the union of $O(g)$ shortest paths in $G'$.
\end{enumerate}
\end{lemma}
\begin{proof}
The cut graph $C$ of $G$ is a union of $O(g)$ noncrossing shortest paths $\{\gamma_1,\ldots ,\gamma_k\}$. By the optimal substructure property, every subpath of a shortest path is a shortest path.  

(1) Let $P'$ be a shortest path in $G'$. Then $P'$ has a connected intersection with each path $\gamma_i$, which is a shortest path in $G$. Every component of $P'\setminus C$ is a shortest path in $G$. Overall, $P'$ has $O(g)$ subpaths along the the paths $\gamma_i$, consequently $P'\setminus C$ has $O(g)$ components. All $O(g)$ subpaths of $P'$ are shortest paths in $G$. 

(2) Let $P$ be a shortest path in $G$. Then $P'$ has a connected intersection with each path $\gamma_i$, which is a shortest path in both $G$ and $G'$. Furthermore, every component of $P\setminus C$ is a shortest path in both $G$ and $G'$. Overall, $P$ decomposes into $O(g)$ subpaths that are shortest paths in both $G$ and $G'$.
\end{proof}

By \Cref{lem:cut}(1), we have $\diam(G')\leq O(g\cdot \diam(G))$. We can compute recursive shortest paths separators in $G'$ as in \Cref{ssec:Steiner4}; but the very first separator is $C$, which consists of $O(g)$ shortest paths. Given a shortest path $P_{uv}$ in $G$, it is a union of $O(g)$ shortest paths in $G'$ by \Cref{lem:cut}(2), each of which can be traced in the recursive decomposition of $G'$ by \Cref{lem:FL22}. Our construction of portals and Steiner spanners now works for $G'$ as described in \Cref{ssec:Steiner4} but the number of portals (and Steiner points) increases by a factor of $O(g)$. With these modifications, the proof of 
\Cref{lem:AdditiveSpanner} in \Cref{ssec:Steiner4}
and \Cref{thm:SteinerPlanar} in \Cref{ssec:SteinerPlanar} 
go through with a constant $O(g)$ factor increase for any constant genus $g$.

\section{Spanners in Planar Domains} \label{sec:obstacle}

In this section, we prove (item 2 in) \Cref{thm:spanner-polygonal} and \Cref{thm:steiner} using tools we develop in \Cref{sec:SF-treecover} and \Cref{sec:Steiner}. First, we show \Cref{lm:metric-relations}, which implies that planar metrics and polyhedral domain are equivalent. 

\MetricRelations*
\begin{proof}
Observe that $\trees \not\cong \planar$ since (the metric induced by) the unweighted cycle graph $C_n$ with $n$ vertices cannot be embedded isometrically into any tree metrics. Thus,  $\trees \sqsubset \planar$. 

Abam, de Berg, and Seraji~\cite{ABS19} showed that $\polydom \sqsubseteq \terrain$ by controlling the elevation of polyhedral terrains. Next, we show that $\terrain\sqsubseteq  \planar$. 

Let $P$ be a set of points on the polyhedral terrain.  Our goal is to show that there exists a planar metric induced by an edge-weighted planar graph $G$ and a subset of points $Q \subseteq V(G)$ such that there exists an isometry from $P$ to $Q$.  

We start by taking the arrangements of geodesic paths between all points in $P$; that is, we draw the shortest path between any two points in $P$ on the terrain. We say that two lines that meet at a point $a$ are  \EMPH{intersecting} if they are not equal on one of the two sides of $a$ as shown in \Cref{fig:linesIntersect}.

\begin{figure}[h!]
    \centering
    \includegraphics[width=.5\textwidth]{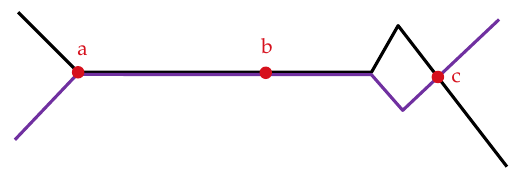}
    \caption{The black and purple lines intersect in $a$ and $c$ but not in $b$}
    \label{fig:linesIntersect}
\end{figure}

We can modify paths so that any two of them only intersect at at most two points. Suppose two shortest paths $p$ and $q$ intersect at three points $a,b$ and $c$. By the suboptimality principle, the section of $p$ and $q$ between $a$ and $b$ are the same length. We modify $q$ to be equal to $p$ between $a$ and $b$. This does not change the length of $q$. After doing the same between $b$ and $c$, $p$ and $q$ only intersect in $a$ and $c$.

We choose an ordering $\sigma: P^2 \mapsto \left[\binom{|P|}{2}\right]$ of the paths and modify the path as described above such that, if two distinct paths $p$ and $q$ with $\sigma(p) < \sigma(q)$ intersect in more than two points, we replace the sections of $q$ with the sections of $p$. This gives an arrangement of geodesic paths such that any two paths intersect at at most two points.

We add Steiner points at all intersection points of the geodesic paths and denote by $G$ the graph obtained. Then $G$ is planar by construction since no pair of edges are intersecting. The point set $Q$ contains the vertices corresponding to $P$. Since all the shortest distances between points in $P$ are preserved in $G$, the natural mapping from every point in $P$ to its copy in $Q$ is an isometry, showing that  $\terrain\sqsubseteq  \planar$.

Next, we show that  $\planar\sqsubseteq  \polydom$. Let $G$ be an edge-weighted planar graph realizing a planar metric $\cM_1$ and $P$ be a subset of vertices of $G$. We draw $G$ on the plane so that every edge is a polygonal curve, with its length being the weight of the edge. Let $R$ be a sufficiently big rectangle that encloses the drawing of $G$. Removing all points (in the drawing) of $G$ from $R$, we get a set of polygonal regions, where each region corresponds to a face of $G$, except one region which corresponds to the intersection of the infinite face of $G$ and $R$.  We now regard each polygonal region as a hole, a.k.a., a polygonal obstacle, thereby obtaining a polygonal domain $\cM_2$. Let $Q$ be the points in $\mathbb{R}^2$ that correspond to the vertices of $P$ in the drawing.  Clearly, the geodesic distance between any two points of $Q$ in $\cM_2$ is the shortest path distance in $G$, and hence (the metric induced by) $P$ can be embedded isometrically int $\cM_2$. 

Lastly, we show that  $\terrain \sqsubset \polysurf$. To show $\polydom \sqsubseteq \polysurf$, consider a point set $P$ on a polyhedral surface, which is a piece-wise linear function $f: D\rightarrow \mathbb{R}$  for some convex polygonal region $D\subset \mathbb{R}^2$. Since $P$ is finite, all shortest paths between points in $P$ lie in a compact subset of the terrains. Consequently, we may assume w.l.o.g.\ that $D$ is compact and (after scaling) lies in a unit square, $D\subset [0,1]^2$; and $f> 0$. Extend $f$ to a larger square domain $[-1,2]^2\subset \mathbb{R}^2$, and let consider the solid $S=\{(x,y,z)\in \mathbb{R}^3: (x,y)\in D \mbox{ and } 0\leq z\leq f(x,y)\}$. Now the boundary of $S$ is a polyhedral surface that contains the terrain (as well as all points in $P$), and the shortest paths among $P$ are the same in both metrics. 

To see that $\terrain\not\sqsubseteq \polysurf$, let $\bM_1$ be the shortest path metric of $K_{3,3}$ (with unit edge weights). It is in $\polysurf$, as $K_{3,3}$ can be realized in $\mathbb{R}^3$ with noncrossing polygonal arcs of unit length, which are shortest paths in some polyhedral surface of sufficiently high genus. However, in any realization of $K_{3,3}$ on a polyhedral terrain $\bM_2$, two shortest paths will cross. Assume that the $a_1b_1$- and $a_2b_2$-paths cross. Then $\delta_2(a_1,a_2)+\delta(b_1,b_2)\leq \delta_2(a_1,b_1)+\delta_2(a_2,b_2) = 2$, in the metric $\delta_2$ of the terrain, and so $\delta_2(a_1,a_2)<2$ and $\delta(b_1,b_2)<2$, while
$\delta_1(a_1,a_2)=\delta_1(b_1,b_2)=2$ in the metric of $K_{3,3}$.
\end{proof}

We are now ready to construct the spanner as claimed in the second item in \Cref{thm:spanner-polygonal} and \Cref{thm:steiner}.

\TerrainSpanner*
\begin{proof}[Proof of item 2] 
    By \Cref{lm:metric-relations}, it suffices to construct a $(2+4\eps)$-spanner for a set of point $P$ in a planar metric realized by an edge-weighted planar graph $G$. One can obtain a $(2+\eps)$-spanner by simply scaling $\eps$. We first construct a tree cover $\mathcal{T}$ for $G$ with $O(\eps^{-3}\log(\eps^{-1}))$ trees using \Cref{thm:tree-cover-planar}. The union of trees of $T$ is a $(1+\eps)$-spanner of $G$. In each tree in $\mathcal{T}$, we remove the Steiner points using \Cref{thm:tree-cover}. This gives a family of $O(\eps^{-3}\log(\eps^{-1}))$ trees on $P$, each with $O(\eps^{-2}\log(\eps^{-1}))$ vertices. The Steiner points removal operation increases the stretch by a factor of $2+\eps$ or less. The union of trees is therefore a non-Steiner spanner for the shortest path metric over $P$, with at most $(1+\eps)(2+\eps) \le (2+4\eps)$ stretch and $O(n\eps^{-5}\log^2(\eps^{-1}))$ edges.
\end{proof}

\SteinerSpanner*

\begin{proof}
By \Cref{lm:metric-relations}, it suffices to construct a Steiner  $(1+\eps)$-spanner for a set of point $P$ of $n$ points in a planar metric realized by an edge-weighted planar graph $G$. Here, we apply   \Cref{thm:SteinerPlanar} on $G$ with $P$ as our set terminals. The number of edges of the spanner is $O((n/\eps)\cdot \log(\eps^{-1}\alpha(n))\cdot \log \eps^{-1})$.
\end{proof}

\section{Lower Bounds}\label{sec:lowerbounds}

\subsection{Stretch 2 Tree Cover}\label{subsec:stretch2-treecover}

We now prove that $\Omega(\log n)$ trees are sometimes necessary  in any non-Steiner tree cover with stretch 2, as claimed in item~2 in \Cref{thm:tree-cover}.  Instead of proving a bound on the number of trees directly, we show a stronger lower bound (\Cref{lm:lb-2-nlogn} below): any $2$-spanner that does not contain any Steiner points must have $\Omega(n\log n)$ edges. As the union of $k$ trees in a Steiner-tree cover with stretch $2$ gives a $2$-spanner with $O(nk)$ edges, it follows that $k = \Omega(\log n)$.

\begin{lemma}\label{lm:lb-2-nlogn} There exists a weighted tree $T$ and a subset of vertices $S \subseteq V(T)$ with $n$ points in $T$ such that any non-Steiner $2$-spanner $G = (S,E, w)$ for $S$ must have $|E| = \Omega(n\log n)$.
\end{lemma}

Let $P_n$ be the unweighted path graph with $n$ vertices $\{1,2,\ldots, n\} = [n]$. We say that an edge-weighted graph $H = ([n], E_H, w_H)$ is a \EMPH{$2$-hop $t$-spanner} for $P_n$ if for every two points $x,y\in P_n$, $d_{P_n}(x,y)\leq d_H(x,y)$ and there exists a path $Q_{xy}$ containing at most \EMPH{two edges} such that $w_H(Q_{x,y}) \leq t\cdot d_{P_n}(x,y)$. In~\cite{LMS22}, the authors showed that any $2$-hop $\frac32$-spanner for $P_n$  must have $\Omega(n\log n)$ edges. Here, we observe that their proof actually implies that any $2$-hop $t$-spanner for any constant $t\geq 1$ must have $\Omega(n\log n)$ edges; we reproduce their poof with the required changes for completeness. Our proof of \Cref{lm:lb-2-nlogn} requires that $t= 2$. 

\begin{lemma}[Adapted from~\cite{LMS22}]\label{lm:2hop-Pn} Let  $H = ([n], E_H, w_H)$ be any $2$-hop $t$-spanner for $P_n$ with $t\geq 1$ and $n\geq 2$, then $|E_H| = \Omega(n\log n/t)$.
\end{lemma}
\begin{proof} We will show a slightly stronger statement by induction. Think of $P_n$ as $n$ integer points on the line, and we allow $H$ to contain edges with integer endpoints outside $P_n$. (But $H$ only needs to preserve distances between points of $P_n$, not the distances to the integer points outside $P_n$.)  Our goal is to lower bound the minimum number of edges, denoted by $T(n)$, in a 2-hop $t$-spanner on $[n]$ with \emph{both endpoints in $P_n$}.

\begin{figure}[h!]
    \centering
    \includegraphics[width=0.9\textwidth]{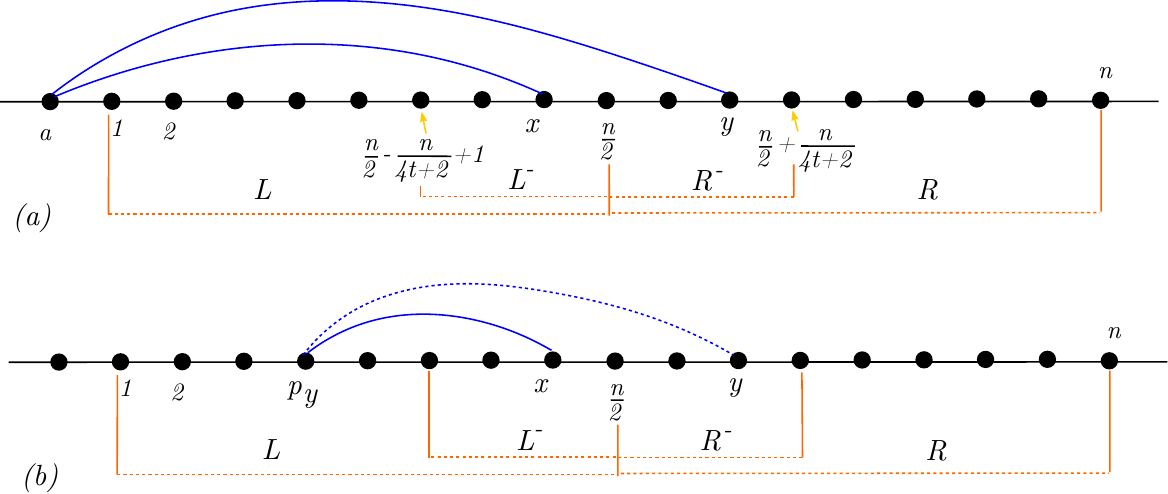}
    \caption{(a) if $Q_{xy}$ contains an integer point $a\leq 0$, then $w(Q_{xy}) > 2 |xy|$; (b) A cross edge $(p_y,y)$ if $x$ is not incident to a cross edge.}
    \label{fig:path-2}
\end{figure}

Let $L = \{1,\ldots, \lfloor n/2\rfloor\}$ be the left half of $P_n$ and $R = P_n\setminus L$ be the right half; see \Cref{fig:path-2}. Note that $E_H$ is a $2$-hop $t$-spanner for both $L$ and $R$.  By induction, $E_H$ has at least $T(\lfloor n/2\rfloor)$ edges with both endpoints in $L$ and $E_H$ has at least $T(|R|)\geq T(\lfloor n/2\rfloor)$ edges with both endpoints in $R$. Note that these two sets are disjoint. Let $E_{C}$ be the set of edges with exactly one endpoint in $L$ and another in $R$, called \EMPH{cross edges}. Then we have 
\begin{equation}\label{eq:Tn}
    T(n) \geq  2  T(\lfloor n/2\rfloor) + |E_C|.
\end{equation}

We now bound the number of cross edges by $\Omega(n/t)$. To avoid notational clutter, we will drop the floors and ceilings, and assume that $n$ is even. Let $L^- = \{n/2-n/(4t+2)+1, n/2-n/(4t+2)+2,\ldots, n/2\}$ be the last $n/(4t+2)$ points of $L$ and similarly, $R^- = \{n/2+1, n/2+2, \ldots, n/2 + n/(4t+2)\}$ be the first $n/(4t+2)$  of $R$. 

We claim that for any two points $x\in L^-$ and $y\in R^-$, any $t$-spanner path $Q_{xy}$ in $H$ of the pair $(x,y)$ must contain only integer points in $P_n$. Since otherwise, w.l.o.g, assume that $Q_{xy}$ contains a point $a \leq 0$, see \Cref{fig:path-2}(a), then
\begin{equation}
    \frac{w(Q_{xy})}{|xy|} >  \frac{n/2-n/(4t+2)} {|xy|} = \frac{n/2-n/(4t+2)}{2 n/(4t+2)} = t,
\end{equation}
contradicting that $Q_{xy}$ is a $t$-spanner path. 

Now we continue to bound $|E_C|$. If every vertex in $L^-$ is incident to a cross edge, which is an edge with an endpoint in $R$, then $|E_C|\geq |L^{-}| = n/(4t+2)$. Otherwise, there is a point $x\in L^-$ that has no crossing edge. For any $y\in R^{-}$, there must be a 2-hop path between $x$ and $y$ of the form $\{x, p_y, y\}$ for some $p_y$ in $L \setminus L^-$ (because the 2-hop path does not leave $P_n$ as claimed above). See \Cref{fig:path-2}(b). Thus, there is a cross edge $(p_y,y)$ for every $y\in R^{-}$, meaning that the number of cross edges is $|E_C|\geq |R^{-}| = n/(4t+2)$.  In both cases, we have $|E_C|\geq   n/(4t+2)$  and \Cref{eq:Tn} becomes
\begin{equation}\label{eq:recurse}
    T(n)\geq 2 T(\lfloor n/2\rfloor) + \Omega(n/t)~.
\end{equation} 
The recurrence in \Cref{eq:recurse} solves to $T(n)= \Omega(\frac{n}{t}\log n)$. 
\end{proof}

\begin{proof}[Proof of \Cref{lm:lb-2-nlogn}] Our lower bound is obtained by a reduction to a lower bound for 2-hop $2$-spanner for the path graph. We construct a tree $T$ as follows. Let $P_n$ be the path with $n$ vertices, where every edge has weight $1$. Let $T$ be obtained by attaching to each vertex $i\in P_n$ a distinct vertex $s_i$ via an edge of weight $w_T(s_i,i) = M$ for an integer $M\geq n$. See \Cref{fig:tree-2}(a) and (b). We call the resulting graph the \EMPH{comb graph}, denoted by $\comb_n$.

Let $S = \{s_1,\ldots, s_n\}$. Note that $d_T(s_i,s_j) = 2M + |j-i|$ for any $j\not=j$, and in particular, $d_T(s_i,s_j)\geq 2M$. Let $G = (S,E,w)$ be any non-Steiner 2-spanner for $S$.

\begin{figure}[h!]
    \centering
    \includegraphics[width=0.8\textwidth]{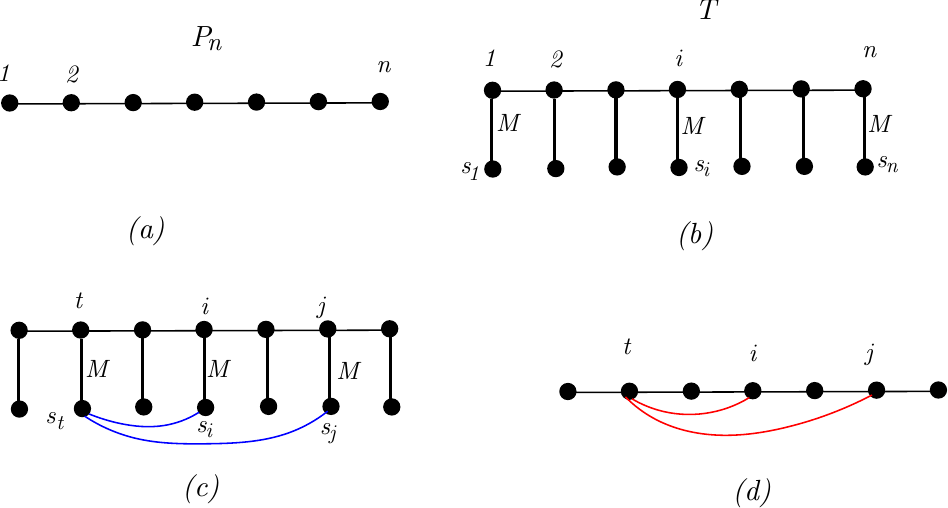}
    \caption{(a) path $P_n$  and (b) tree $T$. Here $M\geq n$. (c) A path $Q_{ij}$ in $G$ between $s_i$ and $s_j$ containing blue edges and (d) the corresponding 2-hop path $Z_{ij}$ in $H$ between $i$ and $j$ containing red edges.}
    \label{fig:tree-2}
\end{figure}

We construct another graph $H = ([n], E_H, w_H)$ as follows: for every edge $(s_i,s_j)$ in $G$, we add an edge $(i,j)$ to $H$ of weight $w_H = |j-i|$. We claim that $H$ is a 2-hop 2-spanner of $P_n$; this will imply that $|E| \geq |E_H| = \Omega(n\log n)$ by \Cref{lm:2hop-Pn}.

Let $i,j$ be any two vertices in $H$ with $j > i$, and two corresponding vertices $s_i$ and $s_j$ in $G$, respectively. If there is an edge $(s_i,s_j) \in G$, then $(i,j) \in E_H$ and hence $(i,j)$ has a path (which is an edge) of stretch $1$ in $H$. 

We now assume that there is no direct edge from $i$ to $j$. Let $Q_{ij}$ be the shortest path from $s_i$ to $s_j$ in $G$. Note that $d_T(s_i,s_j) = 2M +  (j-i)$. Since $G$ is a $2$-spanner of $S$, we have
\begin{equation}\label{eq:dG-ij}
 w(Q_{ij}) \leq 4M + 2(j-i).
\end{equation}

First, we claim that  $Q_{ij}$ contains exactly one point $s_t$ with $t\not\in \{i,j\}$. Suppose otherwise, $Q_{i,j}$ contains at least two points $s_{t_1}, s_{t_2}$. Assume w.l.o.g.\ that $t_1$ is closer to $i$ than $t_2$ on $Q_{ij}$.Then  we have
\begin{equation*}
\begin{split}
    w(Q_{ij}) &\geq d_T(s_i, s_{t_1}) + d_T(s_{t_1}, s_{t_2}) + d_T(t_2, s_{j})\\
    &\geq 2M + 2M + 2M  = 6M >  4M + 2(j-i),    
\end{split}
\end{equation*}
since $M\geq n$, contradicting \Cref{eq:dG-ij}. Thus,  $Q_{ij} = (s_i, s_t,s_j)$ for some $t\neq i,j$; see \Cref{fig:tree-2}(c). Then $w(Q_{ij}) = 4M  + |i-t| + |j-t|$. By \Cref{eq:dG-ij}, we have:
\begin{equation}\label{eq:dH-ij}
    |i-t| + |j-t| \leq  2(j-i).
\end{equation}

Let $Z_{ij} = \{i,t,j\}$; see \Cref{fig:tree-2}(d). Clearly, $Z_{ij}$ is a 2-hop path in $H$ by the construction of $H$. Furthermore, $w_H(Z_{ij}) = |i-t| + |j-t|$ and hence by \Cref{eq:dH-ij}, $Z_{ij}$ has stretch $2$. Thus, $H$ is a $2$-hop $2$-spanner of $P_n$ as claimed.
\end{proof}

\subsection{Spanners in Polyhedral Terrains with Stretch 2}\label{subsec:poly-lb-stretch2}

We now show the lower bound for stretch 2 as claimed in item 1 in \Cref{thm:spanner-polygonal}. 
\TerrainSpanner*
\begin{proof}[Proof of item 1] By \Cref{lm:metric-relations}, any lower bound for non-Steiner spanners in tree metrics will imply the same lower bound for point sets in a polyhedral terrain. Thus, by \Cref{lm:lb-2-nlogn}, there exists a set of $n$ points $P$ such that any $2$-spanner for $P$ must have $\Omega(n\log n)$ edges.
\end{proof}

\subsection{Locality Sensitive Ordering}

In this section, we prove an $\Omega(\log n)$ lower bound on the size of the left-sided LSO stated in \Cref{thm:left-LSo}. We will use 
\LeftLSO*
\begin{proof}
    Let $\comb_n$ be the comb graph constructed in the proof of \Cref{lm:lb-2-nlogn} with terminal set $S$ to be the leaves of $\comb_n$. Let $\Sigma$ be a $(\tau,1)$-left-sided LSO for $\comb_n$. For any linear ordering $\sigma\in \Sigma$, let $\sigma_S$ be the linear ordering obtained by removing all vertices not in $S$ from $\sigma$. Thus, $\sigma_S$ only contains (a subset of) vertices in $S$. Let $\Sigma_S = \{\sigma_S | \sigma \in \Sigma\}$ be the resulting set of linear orderings. 

    We now construct a non-Steiner $2$-spanner for $S$, denoted by $G$, as follows. For each ordering $\sigma_S \in \Sigma_S$, let $v^*_{\sigma_S}$ be the leftmost endpoint of $\sigma_S$. That is, $v^*_{\sigma_S}$ is the smallest vertex in the ordering $\sigma_S$. Then for every $x\in \sigma_S$, we add an edge $(v^*_{\sigma_S}, x)$ to $G$. The weight of every edge in $G$ is the distance between its endpoints in $\comb_n$. This completes the construction of $G$.

    We now argue that $G$ is a spanner with stretch 2 for $S$. Let $s_1,s_2$ be any two vertices in $S$. By the definition of left-sided LSO, there exists a linear ordering $\sigma_S\in \Sigma$ containing both $s_1$ and $s_2$ such that for any $x\preceq_{\sigma_S}s_1$ and $y\preceq_{\sigma_S}s_2$, $d_{\comb_n}(x,y)\leq d_{\comb_n}(s_1,s_2)$. Applying this property to $x = v^*_{\sigma_S}$ and $y = s_2$, we have that $d_{\comb_n}(v^*_{\sigma_S},s_2)\leq d_{\comb_n}(s_1,s_2)$. By a symmetric argument, $d_{\comb_n}(v^*_{\sigma_S},s_1)\leq d_{\comb_n}(s_1,s_2)$. Thus, $d_{\comb_n}(v^*_{\sigma_S},s_1) + d_{\comb_n}(v^*_{\sigma_S},s_2) \leq 2\cdot d_{\comb_n}(s_1,s_2)$, giving that $d_{G}(s_1,s_2)\leq 2\cdot d_{\comb_n}(s_1,s_2)$ since $G$ contains both $(v^*_{\sigma_S},s_1)$ and $(v^*_{\sigma_S},s_2)$. 
    
    Observe that for every ordering $\sigma_S$, we add at most $|\sigma_S|-1$ edges to $G$, where $|\sigma_S|$ is the number of points in $\sigma_S$. Thus, on average, we add at most $1$ edge per vertex to $G$. This means the total number of edges we add to $G$ is at most $\tau n$ since every vertex appears in at most $\tau$ linear orderings in $\Sigma$. Thus, by \Cref{lm:lb-2-nlogn}, $\tau n = \Omega(n\log n)$, implying that $\tau = \Omega(\log n)$ as claimed.
\end{proof}

\subsection{Reliable Spanners}

To show a lower bound for reliable spanners in trees, we use a \emph{density-sensitive} version of \Cref{lm:2hop-Pn} developed in~\cite{LMS23}. Let $P$ be a point set in an interval $[0,L]$ for some $L \geq 1$. We say that $P$ satisfies the \EMPH{unit interval condition} if every unit sub-interval of $[0,L]$ contains \EMPH{at most one} point in $P$.  Le, Milenkovi\'{c}, and Solomonn~\cite{LMS23} showed the following.

\begin{lemma}[Adapted from Lemma 12~\cite{LMS23}]\label{lm:2hop-Steiner-Pn}  Let P be a set of $n\geq 2$  points in the interval $[0, L]$ satisfying the unit interval condition.  Let  $H$ be any Steiner $2$-hop $t$-spanner for $P$ with $t\geq 1$, then $|E_H| = \Omega(n^2\log n/L)$.
\end{lemma}

Le, Milenkovi\'{c}, and Solomonn~\cite{LMS23} stated their lower bound in \Cref{lm:2hop-Steiner-Pn} for stretch $1+\eps$ where $\eps \in (0,1/4]$. However, following the same modification we made in the proof of \Cref{lm:2hop-Pn}, the lower bound holds for \emph{any constant stretch}. Note that in \Cref{lm:2hop-Steiner-Pn}, we allow $H$ to contain points in $[0,L]$ that are not in $P$. The lower bound in \Cref{lm:2hop-Pn} also applies to Steiner spanners, but we do not need this property for the proof of \Cref{lm:lb-2-nlogn}. However, for proving lower bounds on reliable spanners, we do need the lower bound to hold even for Steiner spanners. We now prove \Cref{thm:reliable-lb} which we restate below.

\ReliableSP*
\begin{proof} Let $\comb_n$ be the comb graph (with $2n$ vertices) and $\mathcal{D}$ be a distribution of oblivious $\nu$-reliable $2$-spanners for $\comb_n$ with $\nu  = 1/3$. Our goal is to show that the exists a graph in the support of $\mathcal{D}$ with $\Omega(n\log n)$ edges. 

Let the attack set $B$ contain all the internal nodes of $\comb_n$. By definition,  $\mathbb{E}_{G\sim \mathcal{D}}[|B^+|] \leq (1+\nu)|B|$. Thus, there exists a graph $G\in \mathcal{D}$ such that $|B^+| \leq (1+\nu)|B| = 4n/3$. Next we will show that $|E(G)| = \Omega(n\log n)$.

Observe that there are at least $2n-4n/3 \geq 2n/3$ (leaf) vertices that are not in $B^+$.  Let this set of leaves be $A$. By definition, $G[V(\comb_n)\setminus B]$ is a $2$-spanner for vertices in $A$. Note that $V(\comb_n)\setminus B$ only contain leaves of $\comb_n$.  

Similar to the proof of \Cref{lm:lb-2-nlogn}, we construct another graph $H = ([n], E_H, w_H)$ as follows: for every edge $(s_i,s_j)$ in $G$ between two leaves $s_i, s_j$ of $\comb_n$, we add an edge $(i,j)$ to $H$ of weight $w_H = |j-i|$. Let $A_H = \{j | s_j \in A\}$ be the set of points corresponding to vertices in $A$.

Since  $G[V(\comb_n)\setminus B]$ is a $2$-spanner for vertices in $A$, $H$ is a Steiner $2$-spanner for $A_H$.  By translation, we can assume that vertices of $H$ are in the interval $[n,2n]$, and thus, any Steiner point in $H$ is inside $[0,3n]$. Since $A_H$ satisfies the unit interval condition with $L = 3n$, by \Cref{lm:2hop-Steiner-Pn}, we have:
\begin{equation}
    |E(H)| = \Omega\left(\frac{|A_H|^2\log(|A_H|)}{3n}\right) = \Omega(n\log n)
\end{equation}
since $|A_H| = |A|\geq 2n/3$.  This implies $|E(G)| = \Omega(n\log n)$ as desired.
\end{proof}

\paragraph{Acknowledgements.~} 
This research project started at the \emph{Focused Week: Geometric Spanners}, held at the the Erd\H{o}s Center in Budapest, Hungary, October 23-29, 2023, supported in part by the ERC grant~882971 ``GeoScape''. Balázs Keszegh was supported by the J\'anos Bolyai Research Scholarship of the Hungarian Academy of Sciences, by the National Research, Development and Innovation Office -- NKFIH under the grant K~132696 and FK 132060, by the \'UNKP-23-5 New National Excellence Program of the Ministry for Innovation and Technology from the source of the National Research, Development and Innovation Fund, and by the ERC Advanced Grant ``ERMiD'' and by the Ministry of Innovation and Technology of Hungary from the National Research, Development and Innovation Fund, financed under the ELTE TKP 2021-NKTA-62 funding scheme. 
Hung Le  was supported by the NSF CAREER award CCF-2237288 and the NSF grant CCF-2121952.
Dömötör Pálvölgyi was supported by the J\'anos Bolyai Research Scholarship of the Hungarian Academy of Sciences, by the \'UNKP-23-5 New National Excellence Program of the Ministry for Innovation and Technology, and by the ERC Advanced Grant ``ERMiD''. 
Research by T\'oth was supported in part by the NSF award DMS-2154347.

\small
\bibliographystyle{alphaurl}
\bibliography{ref}

\newcommand{\etalchar}[1]{$^{#1}$}
\begin{thebibliography}{CdVMdM21}

\bibitem[AAHA15]{AAHA15}
Mohammad~Ali Abam, Marjan Adeli, Hamid Homapour, and Pooya~Zafar Asadollahpoor.
\newblock Geometric spanners for points inside a polygonal domain.
\newblock In {\em Proc.\ 31st Symposium on Computational Geometry (SoCG)},
  volume~34 of {\em LIPIcs}, pages 186--197. Schloss Dagstuhl, 2015.
\newblock \href {https://doi.org/10.4230/LIPIcs.SOCG.2015.186}
  {\path{doi:10.4230/LIPIcs.SOCG.2015.186}}.

\bibitem[ACC{\etalchar{+}}96]{ACCDSZ}
Srinivasa Arikati, Danny~Z. Chen, L.~Paul Chew, Gautam Das, Michiel Smid, and
  Christos~D. Zaroliagis.
\newblock Planar spanners and approximate shortest path queries among obstacles
  in the plane.
\newblock In {\em Proc. 4th European Symposium on Algorithms ({ESA})}, volume
  1136 of {\em LNCS}, pages 514--528. Springer, 1996.
\newblock \href {https://doi.org/10.1007/3-540-61680-2_79}
  {\path{doi:10.1007/3-540-61680-2_79}}.

\bibitem[AdBS19]{ABS19}
Mohammad~Ali Abam, Mark de~Berg, and Mohammad Javad~Rezaei Seraji.
\newblock Geodesic spanners for points on a polyhedral terrain.
\newblock {\em {SIAM} Journal on Computing}, 48(6):1796--1810, 2019.
\newblock \href {https://doi.org/10.1137/18m119358x}
  {\path{doi:10.1137/18m119358x}}.

\bibitem[ADD{\etalchar{+}}93]{ADDJS93}
Ingo Alth{\"{o}}fer, Gautam Das, David~P. Dobkin, Deborah Joseph, and
  Jos{\'{e}} Soares.
\newblock On sparse spanners of weighted graphs.
\newblock {\em Discrete Computational Geometry}, 9(1):81--100, 1993.
\newblock \href {https://doi.org/10.1007/BF02189308}
  {\path{doi:10.1007/BF02189308}}.

\bibitem[AFH{\etalchar{+}}04]{AFHKTT04}
Aaron Archer, Jittat Fakcharoenphol, Chris Harrelson, Robert Krauthgamer, Kunal
  Talwar, and \'{E}va Tardos.
\newblock Approximate classification via earthmover metrics.
\newblock In {\em Proc.\ 15th ACM-SIAM Symposium on Discrete Algorithms
  {(SODA)}}, pages 1079--1087, 2004.
\newblock URL: \url{http://dl.acm.org/citation.cfm?id=982792.982952}.

\bibitem[AG06]{AbrahamG06}
Ittai Abraham and Cyril Gavoille.
\newblock Object location using path separators.
\newblock In {\em Proc.\ 25th {ACM} Symposium on Principles of Distributed
  Computing ({PODC})}, pages 188--197, 2006.
\newblock \href {https://doi.org/10.1145/1146381.1146411}
  {\path{doi:10.1145/1146381.1146411}}.

\bibitem[AGMW10]{AbrahamGMW10}
Ittai Abraham, Cyril Gavoille, Dahlia Malkhi, and Udi Wieder.
\newblock Strong-diameter decompositions of minor free graphs.
\newblock {\em Theory Comput. Syst.}, 47(4):837--855, 2010.
\newblock \href {https://doi.org/10.1007/S00224-010-9283-6}
  {\path{doi:10.1007/S00224-010-9283-6}}.

\bibitem[AP90a]{AwerbuchP90b}
Baruch Awerbuch and David Peleg.
\newblock Network synchronization with polylogarithmic overhead.
\newblock In {\em Proc. 31st IEEE Symposium on Foundations of Computer Science
  ({FOCS})}, pages 514--522, 1990.
\newblock \href {https://doi.org/10.1109/FSCS.1990.89572}
  {\path{doi:10.1109/FSCS.1990.89572}}.

\bibitem[AP90b]{AwerbuchP90a}
Baruch Awerbuch and David Peleg.
\newblock Sparse partitions (extended abstract).
\newblock In {\em Proc. 31st IEEE Symposium on Foundations of Computer Science
  ({FOCS})}, pages 503--513, 1990.
\newblock \href {https://doi.org/10.1109/FSCS.1990.89571}
  {\path{doi:10.1109/FSCS.1990.89571}}.

\bibitem[AS87]{AS87}
Noga Alon and Baruch Schieber.
\newblock Optimal preprocessing for answering on-line product queries.
\newblock Technical report, The Moise and Frida Eskenasy Institute of Computer
  Science, Tel Aviv University, 1987.

\bibitem[BDHM07]{BateniDHM07}
MohammadHossein Bateni, Erik~D. Demaine, MohammadTaghi Hajiaghayi, and Mohammad
  Moharrami.
\newblock Plane embeddings of planar graph metrics.
\newblock {\em Discret. Comput. Geom.}, 38(3):615--637, 2007.
\newblock \href {https://doi.org/10.1007/S00454-007-1353-4}
  {\path{doi:10.1007/S00454-007-1353-4}}.

\bibitem[BDMS13]{BDMS13}
Prosenjit Bose, Vida Dujmovi\'c, Pat Morin, and Michiel Smid.
\newblock Robust geometric spanners.
\newblock {\em SIAM Journal on Computing}, 42(4):1720–1736, 2013.
\newblock \href {https://doi.org/10.1137/120874473}
  {\path{doi:10.1137/120874473}}.

\bibitem[BHO20]{BuchinHO20}
Kevin Buchin, Sariel Har{-}Peled, and D{\'{a}}niel Ol{\'{a}}h.
\newblock A spanner for the day after.
\newblock {\em Discret. Comput. Geom.}, 64(4):1167--1191, 2020.
\newblock \href {https://doi.org/10.1007/S00454-020-00228-6}
  {\path{doi:10.1007/S00454-020-00228-6}}.

\bibitem[BHO22]{BuchinHO22}
Kevin Buchin, Sariel Har{-}Peled, and D{\'{a}}niel Ol{\'{a}}h.
\newblock Sometimes reliable spanners of almost linear size.
\newblock {\em J. Comput. Geom.}, 13(1):178--196, 2022.
\newblock \href {https://doi.org/10.20382/JOCG.V13I1A6}
  {\path{doi:10.20382/JOCG.V13I1A6}}.

\bibitem[Bil22]{Bilo22a}
Davide Bil{\`{o}}.
\newblock Almost optimal algorithms for diameter-optimally augmenting trees.
\newblock {\em Theor. Comput. Sci.}, 931:31--48, 2022.
\newblock \href {https://doi.org/10.1016/J.TCS.2022.07.028}
  {\path{doi:10.1016/J.TCS.2022.07.028}}.

\bibitem[BLT14]{BuschLT14}
Costas Busch, Ryan LaFortune, and Srikanta Tirthapura.
\newblock Sparse covers for planar graphs and graphs that exclude a fixed
  minor.
\newblock {\em Algorithmica}, 69(3):658--684, 2014.
\newblock \href {https://doi.org/10.1007/S00453-013-9757-4}
  {\path{doi:10.1007/S00453-013-9757-4}}.

\bibitem[BT22]{BT22}
Sujoy Bhore and Csaba~D. T{\'{o}}th.
\newblock Euclidean {S}teiner spanners: {L}ight and sparse.
\newblock {\em {SIAM} Journal on Discrete Mathematics}, 36(3):2411--2444, 2022.
\newblock \href {https://doi.org/10.1137/22m1502707}
  {\path{doi:10.1137/22m1502707}}.

\bibitem[BTS94]{BodlaenderTS94}
Hans~L. Bodlaender, Gerard Tel, and Nicola Santoro.
\newblock Trade-offs in non-reversing diameter.
\newblock {\em Nord. J. Comput.}, 1(1):111--134, 1994.

\bibitem[CCL{\etalchar{+}}23]{ChangCLMST23}
Hsien{-}Chih Chang, Jonathan Conroy, Hung Le, Lazar Milenkovic, Shay Solomon,
  and Cuong Than.
\newblock Covering planar metrics (and beyond): {O(1)} trees suffice.
\newblock In {\em Proc. 64th {IEEE} Symposium on Foundations of Computer
  Science ({FOCS})}, pages 2231--2261, 2023.
\newblock \href {https://doi.org/10.1109/FOCS57990.2023.00139}
  {\path{doi:10.1109/FOCS57990.2023.00139}}.

\bibitem[CCL{\etalchar{+}}24]{ChangCLMST24}
Hsien{-}Chih Chang, Jonathan Conroy, Hung Le, Lazar Milenkovic, Shay Solomon,
  and Cuong Than.
\newblock Shortcut partitions in minor-free graphs: {S}teiner point removal,
  distance oracles, tree covers, and more.
\newblock In {\em Proc. 35th ACM-SIAM Symposium on Discrete Algorithms
  ({SODA})}, pages 5300--5331, 2024.
\newblock \href {https://doi.org/10.1137/1.9781611977912.191}
  {\path{doi:10.1137/1.9781611977912.191}}.

\bibitem[CdVMdM21]{CohenAddadVMM21}
Vincent Cohen{-}Addad, {\'{E}}ric~Colin de~Verdi{\`{e}}re, D{\'{a}}niel Marx,
  and Arnaud de~Mesmay.
\newblock Almost tight lower bounds for hard cutting problems in embedded
  graphs.
\newblock {\em J. {ACM}}, 68(4):30:1--30:26, 2021.
\newblock \href {https://doi.org/10.1145/3450704} {\path{doi:10.1145/3450704}}.

\bibitem[CG84]{ChungG84}
Fan R.~K. Chung and Michael~R. Garey.
\newblock Diameter bounds for altered graphs.
\newblock {\em J. Graph Theory}, 8(4):511--534, 1984.
\newblock URL: \url{https://doi.org/10.1002/jgt.3190080408}, \href
  {https://doi.org/10.1002/JGT.3190080408} {\path{doi:10.1002/JGT.3190080408}}.

\bibitem[CGN{\etalchar{+}}06]{CGNRS06}
Chandra Chekuri, Anupam Gupta, Ilan Newman, Yuri Rabinovich, and Alistair
  Sinclair.
\newblock Embedding $k$-outerplanar graphs into $\ell_1$.
\newblock {\em SIAM Journal on Discrete Mathematics}, 20(1):119--136, 2006.
\newblock \href {https://doi.org/10.1137/s0895480102417379}
  {\path{doi:10.1137/s0895480102417379}}.

\bibitem[Cha87]{Chazelle87a}
Bernard Chazelle.
\newblock Computing on a free tree via complexity-preserving mappings.
\newblock {\em Algorithmica}, 2:337--361, 1987.
\newblock \href {https://doi.org/10.1007/BF01840366}
  {\path{doi:10.1007/BF01840366}}.

\bibitem[Che86]{Chew86}
L.~Paul Chew.
\newblock There is a planar graph almost as good as the complete graph.
\newblock In {\em Proc.\ 2nd ACM Symposium on Computational Geometry ({SoCG})},
  pages 169--177, 1986.
\newblock \href {https://doi.org/10.1145/10515.10534}
  {\path{doi:10.1145/10515.10534}}.

\bibitem[Che89]{Chew89}
L.~Paul Chew.
\newblock There are planar graphs almost as good as the complete graph.
\newblock {\em Journal of Computer and System Sciences}, 39(2):205--219, 1989.
\newblock \href {https://doi.org/10.1016/0022-0000(89)90044-5}
  {\path{doi:10.1016/0022-0000(89)90044-5}}.

\bibitem[CHPJ20]{CHJ20}
Timothy~M. Chan, Sariel Har-Peled, and Mitchell Jones.
\newblock On locality-sensitive orderings and their applications.
\newblock {\em SIAM Journal on Computing}, 49(3):583--600, 2020.
\newblock \href {https://doi.org/10.1137/19m1246493}
  {\path{doi:10.1137/19m1246493}}.

\bibitem[CKT22]{CKT22}
Hsien-Chih Chang, Robert Krauthgamer, and Zihan Tan.
\newblock Almost-linear $\epsilon$-emulators for planar graphs.
\newblock In {\em Proc.\ 54th ACM Symposium on Theory of Computing {(STOC)}},
  pages 1311--1324, 2022.
\newblock \href {https://doi.org/10.1145/3519935.3519998}
  {\path{doi:10.1145/3519935.3519998}}.

\bibitem[Cla87]{Clarkson87}
Kenneth Clarkson.
\newblock Approximation algorithms for shortest path motion planning.
\newblock In {\em Proc.\ 19th {ACM} Symposium on Theory of Computing {(STOC)}},
  pages 56--65, 1987.
\newblock \href {https://doi.org/10.1145/28395.28402}
  {\path{doi:10.1145/28395.28402}}.

\bibitem[CXKR06]{CXKR06}
T.-H.~Hubert Chan, Donglin Xia, Goran Konjevod, and Andrea Richa.
\newblock A tight lower bound for the {S}teiner point removal problem on trees.
\newblock In {\em Approximation, Randomization, and Combinatorial Optimization.
  Algorithms and Techniques (APPROX/RANDOM)}, volume 4110 of {\em LNCS}, pages
  70--81. Springer, 2006.
\newblock \href {https://doi.org/10.1007/11830924_9}
  {\path{doi:10.1007/11830924_9}}.

\bibitem[dBOP{\etalchar{+}}24]{dOPSW24}
Sarita de~Berg, Tim Ophelders, Irene Parada, Frank Staals, and Jules Wulms.
\newblock The complexity of geodesic spanners using {S}teiner points, 2024.
\newblock \href {http://arxiv.org/abs/2402.12110} {\path{arXiv:2402.12110}}.

\bibitem[dBvKS]{DVS23}
Sarita de~Berg, Marc van Kreveld, and Frank Staals.
\newblock {The Complexity of Geodesic Spanners}.
\newblock In {\em Proc. 39th Symposium on Computational Geometry ({SoCG})},
  volume 258 of {\em LIPIcs}, pages 16:1--16:16. Schloss Dagstuhl.
\newblock \href {https://doi.org/10.4230/LIPIcs.SoCG.2023.16}
  {\path{doi:10.4230/LIPIcs.SoCG.2023.16}}.

\bibitem[dV17]{ECdV17}
\'Eric~Colin de~Verdi\`ere.
\newblock Computational topology of graphs on surfaces.
\newblock In Jacob~E. Goodman, Joeseph O'Rourke, and Csaba~D. T\'oth, editors,
  {\em Handbook of Discrete and Computational Geometry}. CRC Press, Boca Raton,
  FL, 3rd edition, 2017.
\newblock \href {https://doi.org/10.1201/9781315119601}
  {\path{doi:10.1201/9781315119601}}.

\bibitem[EH04]{EricksonH04}
Jeff Erickson and Sariel Har{-}Peled.
\newblock Optimally cutting a surface into a disk.
\newblock {\em Discret. Comput. Geom.}, 31(1):37--59, 2004.
\newblock \href {https://doi.org/10.1007/S00454-003-2948-Z}
  {\path{doi:10.1007/S00454-003-2948-Z}}.

\bibitem[Epp03]{Eppstein03}
David Eppstein.
\newblock Dynamic generators of topologically embedded graphs.
\newblock In {\em Proc.\ 14th {ACM-SIAM} Symposium on Discrete Algorithms
  ({SODA})}, pages 599--608, 2003.
\newblock URL: \url{http://dl.acm.org/citation.cfm?id=644108.644208}.

\bibitem[Fil20]{Filtser20}
Arnold Filtser.
\newblock Scattering and sparse partitions, and their applications.
\newblock In {\em Proc.\ 47th International Colloquium on Automata, Languages,
  and Programming ({ICALP})}, volume 168 of {\em LIPIcs}, pages 47:1--47:20.
  Schloss Dagstuhl, 2020.
\newblock \href {https://doi.org/10.4230/LIPICS.ICALP.2020.47}
  {\path{doi:10.4230/LIPICS.ICALP.2020.47}}.

\bibitem[FL22a]{FL22B}
Arnold Filtser and Hung Le.
\newblock Locality-sensitive orderings and applications to reliable spanners.
\newblock In {\em Proc.\ 54th ACM Symposium on Theory of Computing {(STOC)}},
  pages 1066--1079, 2022.
\newblock \href {https://doi.org/10.1145/3519935.3520042}
  {\path{doi:10.1145/3519935.3520042}}.

\bibitem[FL22b]{FiltserL22}
Arnold Filtser and Hung Le.
\newblock Low treewidth embeddings of planar and minor-free metrics.
\newblock In {\em Proc.\ 63rd {IEEE} Symposium on Foundations of Computer
  Science ({FOCS})}, pages 1081--1092, 2022.
\newblock \href {https://doi.org/10.1109/FOCS54457.2022.00105}
  {\path{doi:10.1109/FOCS54457.2022.00105}}.

\bibitem[FRT04]{FRT04}
Jittat Fakcharoenphol, Satish Rao, and Kunal Talwar.
\newblock A tight bound on approximating arbitrary metrics by tree metrics.
\newblock {\em Journal of Computer and System Sciences}, 69(3):485--497, 2004.
\newblock \href {https://doi.org/10.1016/j.jcss.2004.04.011}
  {\path{doi:10.1016/j.jcss.2004.04.011}}.

\bibitem[GHT84]{GilbertHT84}
John~R. Gilbert, Joan~P. Hutchinson, and Robert~Endre Tarjan.
\newblock A separator theorem for graphs of bounded genus.
\newblock {\em J. Algorithms}, 5(3):391--407, 1984.
\newblock \href {https://doi.org/10.1016/0196-6774(84)90019-1}
  {\path{doi:10.1016/0196-6774(84)90019-1}}.

\bibitem[Goo95]{Goodrich95}
Michael~T. Goodrich.
\newblock Planar separators and parallel polygon triangulation.
\newblock {\em J. Comput. Syst. Sci.}, 51(3):374--389, 1995.
\newblock \href {https://doi.org/10.1006/JCSS.1995.1076}
  {\path{doi:10.1006/JCSS.1995.1076}}.

\bibitem[Gup01]{Gupta01}
Anupam Gupta.
\newblock Steiner points in tree metrics don't (really) help.
\newblock In {\em Proc. 12th ACM-SIAM Symposium on Discrete Algorithms
  ({SODA})}, pages 220--227, 2001.
\newblock URL: \url{http://dl.acm.org/citation.cfm?id=365411.365448}.

\bibitem[HCC11]{HCC11}
{David Ian} Heywood, {Sarah Catharine} Cornelius, and {S. J.} Carver.
\newblock {\em An Introduction to Geographical Information Systems}.
\newblock Pearson Prentice Hall, 2011.

\bibitem[HPMO23]{HMO23}
Sariel Har-Peled, Manor Mendel, and D\'aniel Ol\'ah.
\newblock Reliable spanners for metric spaces.
\newblock {\em ACM Transactions on Algorithms}, (1):1--27, 2023.
\newblock \href {https://doi.org/10.1145/3563356} {\path{doi:10.1145/3563356}}.

\bibitem[JLN{\etalchar{+}}05]{JiaLNRS05}
Lujun Jia, Guolong Lin, Guevara Noubir, Rajmohan Rajaraman, and Ravi Sundaram.
\newblock Universal approximations for {TSP}, {S}teiner tree, and set cover.
\newblock In {\em Proc.\ 37th {ACM} Symposium on Theory of Computing ({STOC})},
  pages 386--395, 2005.
\newblock \href {https://doi.org/10.1145/1060590.1060649}
  {\path{doi:10.1145/1060590.1060649}}.

\bibitem[Kei88]{Keil88}
J.~Mark Keil.
\newblock Approximating the complete {E}uclidean graph.
\newblock In {\em Proc.\ 1st Scandinavian Workshop on Algorithm Theory
  {(SWAT})}, volume 318 of {\em LNCS}, pages 208--213. Springer, 1988.
\newblock \href {https://doi.org/10.1007/3-540-19487-8_23}
  {\path{doi:10.1007/3-540-19487-8_23}}.

\bibitem[KL09]{KL09}
Sanjiv Kapoor and Xiang-Yang Li.
\newblock Geodesic spanners on polyhedral surfaces.
\newblock In {\em Proc.\ 20th International Symposium on Algorithms and
  Computation ({ISAAC})}, volume 5878 of {\em LNCS}, pages 213--223. Springer,
  2009.
\newblock \href {https://doi.org/10.1007/978-3-642-10631-6_23}
  {\path{doi:10.1007/978-3-642-10631-6_23}}.

\bibitem[Kle02]{Klein02}
Philip~N. Klein.
\newblock Preprocessing an undirected planar network to enable fast approximate
  distance queries.
\newblock In {\em Proc. 13th {ACM-SIAM} Symposium on Discrete Algorithms
  {(SODA)}}, pages 820--827, 2002.
\newblock URL: \url{http://dl.acm.org/citation.cfm?id=545381.545488}.

\bibitem[KMS13]{KleinMS13}
Philip~N. Klein, Shay Mozes, and Christian Sommer.
\newblock Structured recursive separator decompositions for planar graphs in
  linear time.
\newblock In {\em Proc. 45th ACM Symposium on Theory of Computing Conference
  ({STOC})}, pages 505--514, 2013.
\newblock \href {https://doi.org/10.1145/2488608.2488672}
  {\path{doi:10.1145/2488608.2488672}}.

\bibitem[KNZ14]{KNZ14}
Robert Krauthgamer, Huy~L. Nguyen, and Tamar Zondiner.
\newblock Preserving terminal distances using minors.
\newblock {\em SIAM Journal on Discrete Mathematics}, 28(1):127--141, 2014.
\newblock \href {https://doi.org/10.1137/120888843}
  {\path{doi:10.1137/120888843}}.

\bibitem[KPR93]{KleinPR93}
Philip~N. Klein, Serge~A. Plotkin, and Satish Rao.
\newblock Excluded minors, network decomposition, and multicommodity flow.
\newblock In {\em Proc.\ 25th {ACM} Symposium on Theory of Computing {(STOC)}},
  pages 682--690, 1993.
\newblock \href {https://doi.org/10.1145/167088.167261}
  {\path{doi:10.1145/167088.167261}}.

\bibitem[LMS22]{LMS22}
Hung Le, Lazar Milenkovi\'{c}, and Shay Solomon.
\newblock Sparse {E}uclidean spanners with tiny diameter: {A} tight lower
  bound.
\newblock In {\em Proc. 38th Symposium on Computational Geometry (SoCG)}, pages
  54:1--54:15, 2022.
\newblock \href {https://doi.org/10.4230/LIPIcs.SoCG.2022.54}
  {\path{doi:10.4230/LIPIcs.SoCG.2022.54}}.

\bibitem[LMS23]{LMS23}
Hung Le, Lazar Milenkovi\'{c}, and Shay Solomon.
\newblock Sparse {E}uclidean spanners with optimal diameter: {A} general and
  robust lower bound via a concave inverse-{A}ckermann function.
\newblock In {\em Proc. 39th Symposium on Computational Geometry (SoCG)}, pages
  47:1--47:17, 2023.
\newblock \href {https://doi.org/10.4230/LIPICS.SOCG.2023.47}
  {\path{doi:10.4230/LIPICS.SOCG.2023.47}}.

\bibitem[LS22]{LS22}
Hung Le and Shay Solomon.
\newblock Truly optimal {E}uclidean spanners.
\newblock {\em {SIAM} Journal on Computing}, pages FOCS19--135--199, 2022.
\newblock \href {https://doi.org/10.1137/20m1317906}
  {\path{doi:10.1137/20m1317906}}.

\bibitem[LT79]{LT79}
Richard~J. Lipton and Robert~Endre Tarjan.
\newblock A separator theorem for planar graphs.
\newblock {\em SIAM Journal on Applied Mathematics}, 36(2):177--189, 1979.
\newblock \href {https://doi.org/10.1137/0136016} {\path{doi:10.1137/0136016}}.

\bibitem[LWN22]{LW21}
Hung Le and Christian Wulff-Nilsen.
\newblock Optimal approximate distance oracle for planar graphs.
\newblock In {\em Proc. 62nd IEEE Symposium on Foundations of Computer Science
  {(FOCS)}}, pages 363--374, 2022.
\newblock \href {https://doi.org/10.1109/focs52979.2021.00044}
  {\path{doi:10.1109/focs52979.2021.00044}}.

\bibitem[NR03]{NR03}
Ilan Newman and Yuri Rabinovich.
\newblock A lower bound on the distortion of embedding planar metrics into
  {E}uclidean space.
\newblock {\em Discret. Comput. Geom.}, 29(1):77--81, 2003.
\newblock \href {https://doi.org/10.1007/S00454-002-2813-5}
  {\path{doi:10.1007/S00454-002-2813-5}}.

\bibitem[Rao99]{Rao99}
Satish Rao.
\newblock Small distortion and volume preserving embeddings for planar and
  {E}uclidean metrics.
\newblock In {\em Proc.\ 15th ACM Symposium on Computational Geometry (SoCG)},
  pages 300--306, 1999.
\newblock \href {https://doi.org/10.1145/304893.304983}
  {\path{doi:10.1145/304893.304983}}.

\bibitem[RS91]{RS91}
Jim Ruppert and Raimund Seidel.
\newblock Approximating the $d$-dimensional complete {E}uclidean graph.
\newblock In {\em Proc.\ 3rd Canadian Conference on Computational Geometry
  {(CCCG)}}, pages 207--210, 1991.

\bibitem[Sol13]{Solomon13}
Shay Solomon.
\newblock Sparse {E}uclidean spanners with tiny diameter.
\newblock {\em {ACM} Trans. Algorithms}, 9(3):28:1--28:33, 2013.
\newblock \href {https://doi.org/10.1145/2483699.2483708}
  {\path{doi:10.1145/2483699.2483708}}.

\bibitem[Tho97]{Thorup97}
Mikkel Thorup.
\newblock Parallel shortcutting of rooted trees.
\newblock {\em J. Algorithms}, 23(1):139--159, 1997.
\newblock \href {https://doi.org/10.1006/JAGM.1996.0829}
  {\path{doi:10.1006/JAGM.1996.0829}}.

\bibitem[Tho04]{Thorup04}
Mikkel Thorup.
\newblock Compact oracles for reachability and approximate distances in planar
  digraphs.
\newblock {\em J. {ACM}}, 51(6):993--1024, 2004.
\newblock \href {https://doi.org/10.1145/1039488.1039493}
  {\path{doi:10.1145/1039488.1039493}}.

\bibitem[Yao82]{Yao82}
Andrew~Chi{-}Chih Yao.
\newblock Space-time tradeoff for answering range queries (extended abstract).
\newblock In {\em Proc. 14th {ACM} Symposium on Theory of Computing {(STOC)}},
  pages 128--136, 1982.
\newblock \href {https://doi.org/10.1145/800070.802185}
  {\path{doi:10.1145/800070.802185}}.

\end{thebibliography}

\end{document}